\documentclass[letterpaper,twocolumn,10pt]{article}
\usepackage{usenix2019_v3}

\newif\ifblind
\blindtrue

\ifblind

\else

\fi

\def\vulns{four}
\def\devices{14}
\def\chips{16}

\def\Ezero{\ensuremath{E_{0}}}

\def\cone{{\large \textcircled{\footnotesize 1}}}

\def\Kbt{$\mathrm{K}_{\mathrm{BT}}$}
\def\Kble{$\mathrm{K}_{\mathrm{BLE}}$}
\def\SKbt{$\mathrm{SK}_{\mathrm{BT}}$}
\def\SKble{$\mathrm{SK}_{\mathrm{BLE}}$}

\def\kdfbt{kdf$_{\mathrm{BT}}$}
\def\kdfble{kdf$_{\mathrm{LE}}$}

\usepackage{microtype}
\usepackage{graphicx}
\usepackage{amssymb}
\usepackage[cmex10]{amsmath}
\usepackage[tight,footnotesize]{subfigure}
\usepackage{caption}

\usepackage{msc}

\usepackage{booktabs}
\usepackage{enumerate}
\usepackage{hyperref}
\usepackage{wasysym}
\usepackage{fontawesome}
\usepackage{hyperref}

\newcommand{\ie}{i.e.,}
\newcommand{\eg}{e.g.,}

\newcommand{\Xyes}{$\CIRCLE$}
\newcommand{\Xpar}{$\LEFTcircle$}
\newcommand{\Xno}{$\Circle$}
\newcommand{\design}{Standard}
\newcommand{\impl}{Impl.}
\newcommand{\persistent}{\checkmark}
\newcommand{\notpersistent}{-}
\newcommand{\scon}{\checkmark}

\begin{document}

\date{}

\title{\Large \bf BLURtooth: Exploiting Cross-Transport Key Derivation in Bluetooth Classic and Bluetooth Low Energy}

\author{
{\rm Daniele Antonioli}\\ EURECOM \and
{\rm Nils Ole Tippenhauer}\\ CISPA Helmholtz Center \and
{\rm Kasper Rasmussen}\\ University of Oxford \and
{\rm Mathias Payer}\\ EPFL
}

\maketitle


\begin{abstract}

The Bluetooth standard specifies two transports: Bluetooth Classic (BT)
for high-throughput wireless services and Bluetooth Low Energy (BLE) for
very low-power scenarios. BT and BLE have dedicated pairing protocols and
devices have to pair over BT and BLE to use both securely. In 2014, the
Bluetooth standard (v4.2) addressed this usability issue by introducing
\emph{Cross-Transport Key Derivation (CTKD)}. CTKD allows establishing BT and
BLE pairing keys just by pairing over one of the two transports. While CTKD
crosses the security boundary between BT and BLE, little is known
about the internals of CTKD and its security implications.

In this work, we present the first complete description of CTKD obtained by
merging the scattered information from the Bluetooth standard with the results from
our reverse-engineering experiments. Then, we perform a security evaluation of
CTKD and uncover four cross-transport issues in its specification.
We leverage these issues to design
four standard-compliant attacks on CTKD enabling new ways to exploit Bluetooth
(\eg\ exploiting BT and BLE by targeting only one of the two). Our attacks work
even if the strongest security mechanism for BT and BLE are in place,
including Numeric Comparison and Secure Connections. They
allow to impersonate, man-in-the-middle, and establish unintended sessions
with arbitrary devices.
We refer to our attacks as \emph{BLUR attacks}, as they \emph{blur} the security
boundary between BT and BLE. We provide a low-cost implementation of the BLUR
attacks and we successfully evaluate them on \devices\ devices with \chips\
unique Bluetooth chips from popular vendors.
We discuss the attacks' root causes and
present effective countermeasures to fix them. We disclosed our findings and
countermeasures to the Bluetooth SIG in May 2020 (CVE-2020-15802),
and we reported additional unmitigated issues in May 2021.

\end{abstract}


\section{Introduction}
\label{sec:intro}

Bluetooth is a pervasive wireless technology used by billions of devices
including mobile phones, laptops, headphones, cars, speakers, medical, and
industrial appliances~\cite{btsig-markets}. It is specified in an open
standard maintained by the Bluetooth special interest group (SIG), and its
latest version is 5.2~\cite{btsig-52}. The standard specifies two transports:
\emph{Bluetooth Classic (BT)} and \emph{Bluetooth Low Energy (BLE)}. BT is
best suited for connection-oriented and high-throughput use cases, such as streaming audio. BLE is optimized for connection-less and very-low-power use
cases such as fitness tracking.

The Bluetooth standard defines dedicated security architectures and threat
models for BT~\cite[p.~947]{btsig-52} and BLE~\cite[p.~1617]{btsig-52}. Each
transport provides a \emph{pairing} and a \emph{session establishment} protocol.
Pairing results in the establishment of a long-term pairing key that acts as the
root of trust.
Session establishment allows paired devices to establish a secure
channel through a fresh session key derived from their shared pairing key.

Traditionally, two devices supporting BT and BLE would have to pair
separately on each transport. In 2014, the Bluetooth standard (v4.2)
introduced \emph{Cross-Transport Key Derivation (CTKD)} to address this
usability issue. CTKD enables to pair devices once, either over BT or BLE,
and negotiate BT and BLE pairing keys without having to pair a second
time~\cite[p.~1401]{btsig-52}.


CTKD has not received any attention from the research community
and the Bluetooth standard  describes only some aspects and threats
associated with CTKD. We believe CTKD provides a significant  attack surface, as it is a standard-compliant security feature, used in conjunction with the most secure Bluetooth modes (\eg\ Secure Connections),
and is transparent to the end-users.
In addition, CTKD allows \emph{crossing} the security boundary between BT and BLE as CTKD forces implicit trust between BT and BLE.
For example, if two devices pair over BT and generate a BLE pairing key with CTKD then the security of the BLE transport entirely relies on BT.

In this work, we present a complete description of CTKD obtained by combining
the incomplete and scattered information provided in the Bluetooth standard 
and our reverse-engineering experiments needed to understand how
CTKD is negotiated and used in practice for BT and BLE. Then, we perform the
first  security evaluation of CTKD and we uncover four \emph{cross-transport issues (CTI)} with its
specification. The issues affect pairing states, role asymmetries, key
generation and distribution, and association methods. For example, CTKD
by-design enables overwriting trusted keys with malicious ones across
transports.

We leverage the uncovered CTIs to design four \emph{cross-transport}
and \emph{standard-compliant}
attacks on CTKD. Our attacks enable persistent cross-transport impersonation, man-in-the-middle,
and unintended session attacks on BT and BLE via CTKD. The attacks are
effective regardless of the employed Bluetooth security mechanisms,
including Secure Connections and Numeric Comparison.
We name our attacks \emph{BLUR attacks}, as they
blur the
security boundary between BT and BLE.

The BLUR attacks are the first standard-compliant BT and BLE
attacks to not require the attacker to be present when a victim is pairing or establishing a secure session, unlike prior work~\cite{hypponen2007nino, haataja2010two, ryan2013bluetooth, sun2018man,
biham2018breaking, antonioli19knob, antonioli20tops,
antonioli20bias,wu20blesa,zhang2020breaking,von21method}. In particular, our attacks are the first that can be conducted in absence of one of the victims. 
The BLUR attacks are also the first that exploit interactions between BT and BLE (via CTKD).
 For a more detailed comparison to prior work see Section~\ref{sec:related}.

To demonstrate that the BLUR attacks are feasible we present a low-cost
implementation of the attacks based on a cheap development board and
open-source software. We evaluated our attacks on a large and heterogeneous
sample of devices. In particular, we exploited
\emph{\devices} unique devices employing \emph{\chips} different Bluetooth chips from
Broadcom, Cambridge Silicon Radio (CSR), Cypress, Intel, and Qualcomm.
Our set of vulnerable devices covers \emph{all} Bluetooth versions supporting CTKD
(\ie\ Bluetooth 4.2, 5.0, 5.1, and 5.2) and even a 4.1 device to
which CTKD was backported.

We concretely address the BLUR attacks by presenting four
\emph{protocol-level} countermeasures mitigating the presented CTIs and the BLUR attacks.
Our mitigations
can be implemented at the operating system level with low effort. To backup
this claim we tested one countermeasure (\ie\ disable key overwriting) by
implementing it on a Linux laptop.

We responsibly disclosed our findings with the Bluetooth SIG two times. In
May 2020 we sent our first report which was tracked with CVE-2020-15802.
In September 2020 the Bluetooth SIG unilaterally released a security note (see
\url{https://tinyurl.com/vxhwftc2}), claiming that Bluetooth 5.1 and later are not vulnerable to the presented attacks.
As result, we further analyzed 5.1 and 5.2 devices, and found them to still be
susceptible. We explain why this is the case in
Section~\ref{sec:blur-discussion} and experimentally confirm it in
Section~\ref{sec:eval-results}. We disclosed those findings to the SIG, but
have not received a reaction.
We note that the SIG is expected to notify vendors of vulnerabilities, so no separate vendor disclosure is required.

We summarize our main contributions as follows:
\begin{itemize}

\item We present a complete description of CTKD combining public and
    reverse-engineered information. We perform the first security evaluation
    of CTKD and uncover four vulnerabilities in its specification. For example, CTKD enables to adversarially pair over unused transports and
    to tamper with BT and BLE security keys.

\item Based on the identified issues we propose four novel and
    standard-compliant attacks capable of breaking BT and BLE just by
    targeting one of the two. Compared to related work, our attacks are the
    first exploiting CTKD and acting across transports.
    Our attacks enable to impersonate, man-in-the-middle, and
    establish unwanted and stealthy sessions with arbitrary devices. We name
    our attacks \emph{BLUR attacks} as they blur the security boundary between BT and
    BLE.

\item We present a low-cost implementation of the BLUR
  attacks based on a Linux laptop and a Bluetooth development board.
  We use our implementation to attack \devices\ different devices employing \chips\
  unique Bluetooth chips and covering all Bluetooth versions
  compatible with CTKD (\eg\ 4.2, 5.0, 5.1, and 5.2). Our evaluation
  demonstrates that the BLUR attacks are very effective and
  specification-compliant. To address them, we discuss four countermeasures to
  address the presented issues and attacks affecting CTKD.
\end{itemize}


\section{Bluetooth Classic (BT) and Low Energy (BLE)}
\label{sec:background}


BT and BLE are two wireless transports specified in the Bluetooth
standard~\cite{btsig-52}.
These transports are designed to complement each other. BT is used for
high-throughput and connection-oriented services, such as streaming audio and
voice, while BLE is optimized for very low-power and low-throughput services
such as fitness tracking and digital contact tracing. High-end devices, such
as laptops, smartphones, headsets, and tablets, provide both BT and BLE, while
low-end devices such as mice, keyboards, and wearables provide either BT or
BLE.

BT and BLE have similar security mechanisms (\ie\ pairing and session
establishment) but \emph{different} security architectures and threat models.
Pairing, also known as Secure Simple Pairing (SSP), lets two devices establish
and authenticate a pairing key that acts as the root of trust. 
BLE SSP is performed over the
Security Manager Protocol (SMP)~\cite[p.~1666]{btsig-52}, while BT SSP
uses the Link Manager Protocol (LMP)~\cite[p.~568]{btsig-52}.
During pairing, BLE allows negotiating the entropy of the pairing key
while BT does not.

While pairing, BT and BLE employ similar \emph{association mechanisms}.
For example, \emph{Just Works} association is supported by all Bluetooth
devices supporting pairing as it does not require user interaction,
but it does not protect against MitM attacks. While
\emph{Numeric Comparison} association protects against MitM attacks by
asking the user to confirm a numeric code on the pairing devices'
screens. As the Bluetooth standard does not protect the negotiation of the association
method, an attacker can always \emph{downgrade} it to Just Works even if
the victim device has I/O capabilities.

Session establishment
lets paired devices establish a secure communication channel. The channel is
protected by a fresh session key derived from the pairing key and some nonces.
During session establishment, BT allows negotiating the entropy of the session
key while the BLE session key inherits the entropy of the associated pairing
key.

BT and BLE use the same notion of \emph{pairable} and \emph{discoverable} states.
If a device is pairable then it accepts pairing requests from
other devices. If it is discoverable it reveals its identity
when scanned by other devices. Notably,
a device answers to a pairing request even if it
is \emph{not} discoverable~\cite{bt-miscon2}. For example, if the user knows the
Bluetooth address of her pair of headphones she can complete BT or BLE pairing
by sending a pairing request from her laptop without putting the headphones
into discoverable mode.

BT and BLE provide a \emph{Secure Connections} mode which enhances the security
primitives in use without affecting the underlying protocols. In particular, Secure
Connections mandates the usage of FIPS-compliant algorithms such as AES-CCM,
HMAC-SHA-256, and the ECDH on the P-256 curve~\cite[p.~269]{btsig-52}.


Both BT and BLE use a \emph{master-slave} medium access protocol. The master
(BLE central) is the connection initiator, while the slave (BLE peripheral) is
the responder. BT allows to switch roles dynamically, while BLE roles are
fixed. High-end devices, such as laptops and smartphones, support both BLE master and
BLE slave modes and are typically used as BLE masters, while low-end devices,
such as fitness trackers and smartwatches support only the BLE slave mode
\footnote{For precise technical descriptions in the rest of the paper we follow
the \emph{Bluetooth standard}'s master/slave terminology instead of more apt
terms like leader/follower.}.


\section{Description and Security Analysis of CTKD}
\label{sec:ctkd-eval}

In this section, we present the first complete description and  security
analysis of CTKD\@.
In Section~\ref{sec:ctkd} we describe what is publicly known about CTKD, in
Section~\ref{sec:ctkd-re} we complement those information with other
crucial ones that we had to reverse-engineer (\eg\ CTKD negotiation for
BT and BLE). In Section~\ref{sec:ctkd-vulns},
we uncover four critical and novel  \emph{cross-transport issues (CTI)} in
the specification of CTKD.
These issues are the root causes of the standard-compliant and cross-transport
attacks presented in Section~\ref{sec:impl} and  evaluated in
Section~\ref{sec:eval}.

\subsection{Public Information about CTKD}
\label{sec:ctkd}

As described in the Introduction, CTKD was introduced in the Bluetooth
standard to improve the \emph{usability} of BT and BLE pairing. Before
the introduction of CTKD, devices had to separately pair over BT and
BLE. While with CTKD, the devices pair once, either over BT or BLE,
derive a pairing key for each transport and establish BT and BLE secure
sessions~\cite[p.~280]{btsig-52}.

Being a standard-compliant feature
CTKD has to be supported by all hardware and software
Bluetooth vendors. The list of vendors includes Apple~\cite{ctkd-apple},
Google~\cite{ctkd-fluoride}, Cypress~\cite{ctkd-cypress},
Linux~\cite{ctkd-bluez}, Qualcomm~\cite{ctkd-qualcomm}, and
Intel~\cite{ctkd-intel}. Notably, Apple presented it as a core and always-on
Bluetooth feature during WWDC 2019.

To use CTKD, a device requires few capabilities. It must be a
dual-mode device (\ie\ support both BT and BLE), has to support Secure
Connections, and implement a Bluetooth version among
4.2, 5.0, 5.1, and 5.2. Examples of devices supporting CTKD are laptops,
tablets, smartphones, headsets, speakers, and high-end wearable devices. The
number of those devices is steadily growing as dual-mode devices
are replacing single-mode ones~\cite{btsig-market20}.

CTKD employs the same deterministic \emph{key derivation function (KDF)}
for BT and BLE~\cite[p.~1658]{btsig-52}.
The KDF takes as inputs a 128-bit (16-byte) key and two 4-byte strings and derives a
128-bit (16-byte) key. If CTKD is started from BLE, then
the BT pairing key is derived using the ``tmp2'' and ``brle'' strings. In the other
case, the derivation is performed using the ``tmp1'' and ``lebr'' strings.
The key derivation function is deterministic, as using CTKD on the same
input key will always generate the same output key. We re-implemented KDF to
validate our analysis, see Section~\ref{sec:impl-ctkd} for more
details.


The Bluetooth standard lacks a security analysis of CTKD but provides only
a \emph{limited} and \emph{version-specific} security argument. Since
version 5.1 the standard states that ``While performing
cross-transport key derivation, if the key for the other transport already
exists, then the devices shall not overwrite that existing key with a key that
is weaker in either strength or MITM protection''~\cite[p.~1401]{btsig-52}.
In other words, an attacker cannot overwrite a pairing key with CTKD
if the overwriting key has either a lower entropy (\ie\ strength) or a
lower MitM protection. While this can be expected to protect against attacks in limited settings, other scenarios and attacks are still possible. For example,
an adversary can still overwrite keys with \emph{equal} strength and MitM
protection \emph{without} violating the standard (as we experimentally demonstrate in Section~\ref{sec:eval-results}).
In addition to the limited scope of the countermeasure,  it is unclear why it was
introduced only for 5.1 and 5.2 devices and not for all Bluetooth versions
supporting CTKD, and how it should be interpreted when
one of the devices does not support Bluetooth 5.1 or 5.2.



\subsection{Reverse-Engineered CTKD Protocols}
\label{sec:ctkd-re}

The public information that we gathered about CTKD, including the ones
provided by the Bluetooth standard are \emph{not} sufficient to perform a
security analysis of CTKD. Specifically, from the standard is not clear how
CTKD is \emph{negotiated} for BT and BLE and if the protocols differ. To
address this problem, we reverse-engineered the CTKD negotiation protocols for BT and
BLE. Here we present them abstracting the description at the message
level. We refer to the Bluetooth master as Alice and to the slave as Bob and
in the figure we color-code BLE with light blue and BT with blue.
At the end of the section, we detail our RE methodology.

\begin{figure}[tb]
    \centering
    \includegraphics[width=1.0\linewidth]{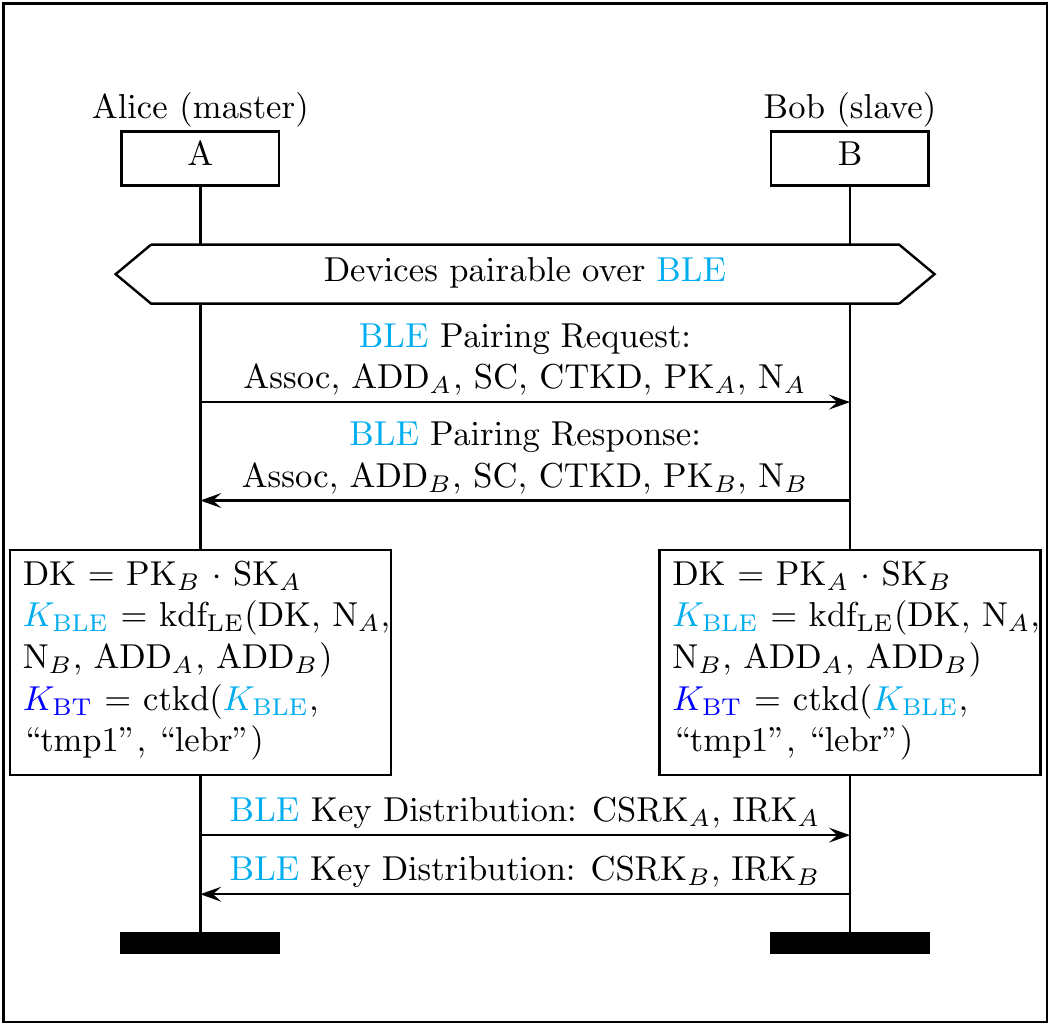}
    \caption{CTKD from BLE. Alice and Bob
    negotiate SC and CTKD support during BLE pairing. Then, they compute the BLE pairing key and from
    that key, they derive the BT pairing key via CTKD (without exchanging any message
    over BT). Finally, they generate and exchange additional keys for BLE
    including signature (CSRK) and identity resolving (IRK) keys. After the
    protocol is completed Alice and Bob can establish
    secure sessions both for BT and BLE (without having to pair over BT).}
    \label{msc:ctkd-ble}
\end{figure}

\textbf{CTKD from BLE} Figure~\ref{msc:ctkd-ble} shows how CTKD is
negotiated and used from BLE to derive BLE and BT pairing keys.
Alice and Bob are pairable over BLE and BT and discover each other using
BLE scanning and advertising. Then, they perform pairing over BLE using the
SMP protocol. We found that CTKD is negotiated by setting to one the Link Key flag
of the Initiator and Responder key distribution SMP fields~\cite[p.~1680]{btsig-52}
and that such negotiation is not protected. Other than the Link Key flag the
devices should also declare Secure Connections
support (SC) which is also spoofable. The BLE pairing messages also contain
an association method (Assoc), a source BLE address (ADD), a public key (PK), and a nonce (N).

After exchanging the pairing messages, the devices compute a Diffie-Hellman
shared secret (DK) using the exchanged PK. DK is used to
compute the BLE pairing key (\Kble) using BLE pairing key derivation function
(\kdfble). Then, the devices use CTKD's key derivation function (ctkd) to
derive the BT pairing key (\Kbt). To complete BLE pairing, Alice and Bob establish a secure
session over BLE and exchange additional keys (\eg\ CSRK and IRK). As a
result, Alice and Bob share \Kble\ and \Kbt, but they only paired over BLE.

\begin{figure}[tb]
    \centering
    \includegraphics[width=1.0\linewidth]{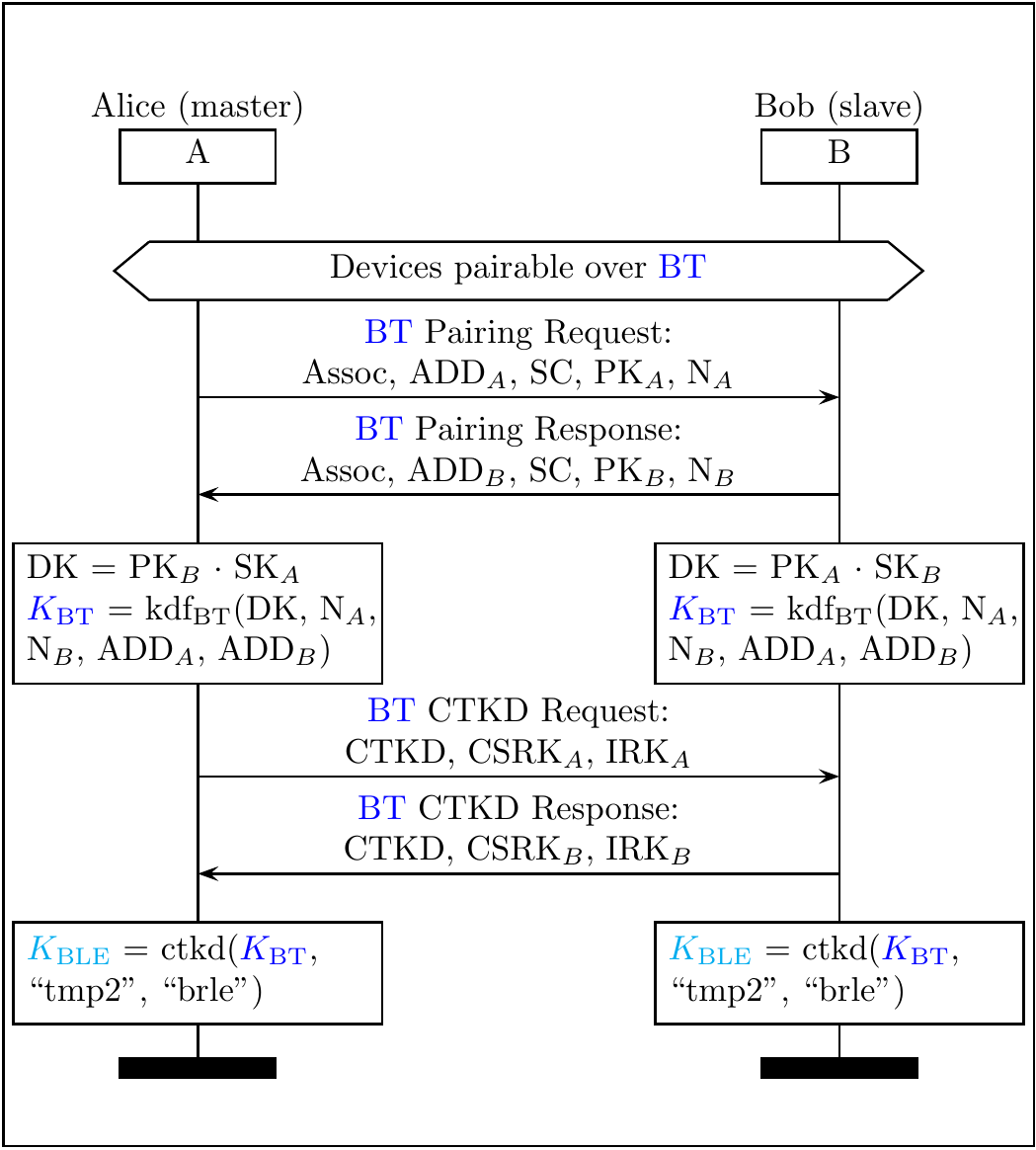}
    \caption{CTKD from BT. Alice and Bob during BT pairing
    negotiate SC support. Then, they compute the BT pairing key, start a
    secure session over BT and send BT CTKD messages containing
    CTKD support and other keying material generated for BLE such as signature
    (CSRK) and identity resolving (IRK) keys. Notably, the CTKD request and
    response are encoded as BLE pairing request and response and tunneled over BT.
    Afterward, Alice and Bob derive the
    BLE pairing key, via CTKD (without exchanging any message over BLE). After the
    protocol is completed Alice and Bob can establish secure sessions both for
    BT and BLE (without having to pair over BLE).}
    \label{msc:ctkd-bt}
\end{figure}

\textbf{CTKD from BT} Figure~\ref{msc:ctkd-bt} presents how CTKD is
negotiated and used from BT to derive BT and BLE pairing keys.
Alice and Bob are pairable over BT and BLE and discover each other
via BT inquiry. Then, they exchange pairing request and response messages over
BT to negotiate several BT capabilities (including SC), and to exchange their
BT addresses, keys, and nonces. Then, they compute DK and use it together with
their BT addresses and nonces to compute the BT pairing key (\Kbt) through the
BT pairing key derivation function (\kdfbt).

Unlike for BLE, BT pairing messages do not include a CTKD flag. What happens
is that the devices start a secure BT session and exchange two messages
containing the CTKD flag and additional
security material needed for BLE such as signature keys (CSRK) and identity
resolving keys (IRK). These two messages are peculiar as they are encoded
as BLE SMP packets but sent over BT. We are not sure why the Bluetooth
standard is not describing such "BLE tunneling" protocol to negotiate CTKD
from BT. Once CTKD is negotiated, Alice and Bob use it to derive the BLE pairing key
(\Kble) from the BT key and the static strings ``tmp2'' and ``brle''.



\textbf{RE methodology} To RE the negotiation and usage of CTKD we used a
Linux laptop connected to a dual-mode development board as a test device. The
laptop runs a patched Linux kernel capable of pairing diagnostic messages from
the board. The board acts as the laptop fronted (\ie\ the laptop is the
BT/BLE Host while the board is the BT/BLE Controller), and is initialized to
report to the laptop all sent and received link-layer
traffic using HCI diagnostic messages.

To test CTKD from BLE we sent a BLE pairing request from our test device to a
pair of dual-mode headphones (Sony WH-1000XM3) and we monitored the HCI log.
To check out CTKD from BT we sent a BT pairing request from our test device to
an Android smartphone (Pixel 2) and we monitored the HCI log. In each case, we
tested that it was possible to establish BT and BLE secure sessions after only
pairing on one transport. Notably, CTKD from BT was
particularly tricky to reverse as the CTKD negotiation messages over BT
are decoded by Wireshark but appear as standard L2CAP messages.



\subsection{CTKD Cross-Transport Issues (CTI)}
\label{sec:ctkd-vulns}

We isolated \vulns\ \emph{cross-transport issues (CTI)} with the specification of
CTKD resulting from CTKD bridging BT and BLE without
properly enforcing the security boundary between the two. We now describe
in detail each CTI.

\textbf{CTI 1: extended pairing} CTKD introduces more options
to pair two devices as dual-mode devices are pairable over BT and BLE all the
time. This enables an attacker to (silently) pair over a transport that is
currently unused. The attacker does not need to wait until a victim is in
discoverable mode, as, despite common belief, a Bluetooth device in
pairable state already accepts pairing requests.

\textbf{CTI 2: role asymmetry} While BT and BLE roles are defined
differently, CTKD does not enforce which role was used to pair on which
transport. BT roles can be switched even before pairing, while BLE roles are
fixed. This is problematic because an attacker can adversarially
switch BT role before using CTKD and send
a BT pairing request to a victim which expects BT and BLE
pairing responses. We note that, issues with role asymmetry have been
already proven effective to bypass BT authentication during session
establishment~\cite{antonioli20bias}.

\textbf{CTI 3: key tampering} CTKD enables to tamper with
all BT security keys from BLE and vice versa using only a single
run of the pairing protocol. This is a new and powerful attack primitive for
Bluetooth. For example, an attacker can use CTKD to write new pairing
keys for BT and BLE or even overwrite trusted pairing keys with her own.
Furthermore, by using CTKD from BT the attacker can get access to all BLE
security keys distributed as part of BLE pairing including identity resolving
key usable to de-anonymize a BLE device.

\textbf{CTI 4: association manipulation} CTKD does not keep track of which
association mechanism was used as part of pairing and the negotiation of the
association scheme is not protected. Indeed, an
attacker can use CTKD to re-establish pairing keys using an arbitrary
association scheme. This includes a weak association to write or substitute
authenticated keys with unauthenticated ones (\eg\ by re-pairing using Just
Works). Recently, association confusion attacks have been proposed for BT or
BLE~\cite{von21method}, CTKD extends this issue across transports.



\section{Attacks via CTKD}
\label{sec:blur}

We now present our threat model and the design of four novel and
standard-compliant attacks on CTKD. Our attacks are the first samples of
\emph{cross-transport} exploitation for Bluetooth, as they are capable of
exploiting BT and BLE just by targeting either of the two. Moreover, they are
the first attacks exploiting CTKD.
The attacks do \emph{not}
require a strong attacker model. For example, they can be conducted
at any time against arbitrary devices (including the ones supporting
BT and BLE SC, and SSP with strong association).
As our attacks are blurring the security boundary between
BT and BLE, we name them \emph{BLUR attacks}.

\subsection{System Model}
\label{sec:sm}

Our system model considers two victims, Alice and Bob, who can securely
communicate over BT and BLE. The victims support CTKD, and are using the most
secure BT and BLE modes, namely, SC and SSP with strong association.
This setup \emph{should} protect the victims
against eavesdropping, impersonation, and
man-in-the-middle attacks as claimed in~\cite[p.~269]{btsig-52}.
Without loss of generality, we assume that Alice is the master
and Bob is the slave.

Regarding the notation, we indicate a BT pairing key with \Kbt, a BT session
key with \SKbt, a BLE pairing key with \Kble, a BLE session key with
\SKble. We indicate a Bluetooth address with ADD,
a public key with PK, a private key with SK, a shared Diffie-Hellman secret
with DK, a nonce with N, and a message authentication code with MAC.




\subsection{Attacker Model and Goals}
\label{sec:am}


Our attacker model considers Charlie, an attacker in Bluetooth
range with the victims.
The attacker's knowledge is limited to what the victims
advertise over the air, \eg\ full or partial Bluetooth addresses, Bluetooth
names, and security and IO capabilities.
She can scan and discover devices, send pairing requests
and responses, use CTKD, propose weak association mechanisms (\eg\ Just Works),
and dissect and craft Bluetooth packets.
However, the attacker does not know any pairing or session key shared between
the victims, and does not have to be present when the victims pair or
negotiate a secure session. Moreover, she cannot access and tamper with the victim
devices.

The attacker has four goals. (i) \emph{impersonate Alice (master)}
and take over her secure sessions with Bob. (ii)
\emph{impersonate Bob (slave)} and take over his secure sessions with Alice.
(iii) \emph{man-in-the-middle} Alice and Bob' secure session
(iv) establish \emph{unintended and stealthy sessions} with Alice and Bob.

Let us clarify some aspects of the attackers' goals. By ``take over'' we mean that
after the attack the bond between the victims is broken (\eg\ when Charlie
takes over a session from Alice then Alice will not be able to connect back
to Bob). Master and slave impersonation are indicated as different goals, as
they require different attack strategies. We define an unintended session as
a session established with a victim device as an arbitrary device without
breaking existing security bonds (\eg\ Charlie silently pairs with Alice as a
random device using CTKD and connects with Alice over BT and BLE).



\subsection{Attack Strategy}
\label{sec:blur-as}

\begin{figure}[tb]
    \centering
    \includegraphics[width=.8\linewidth]{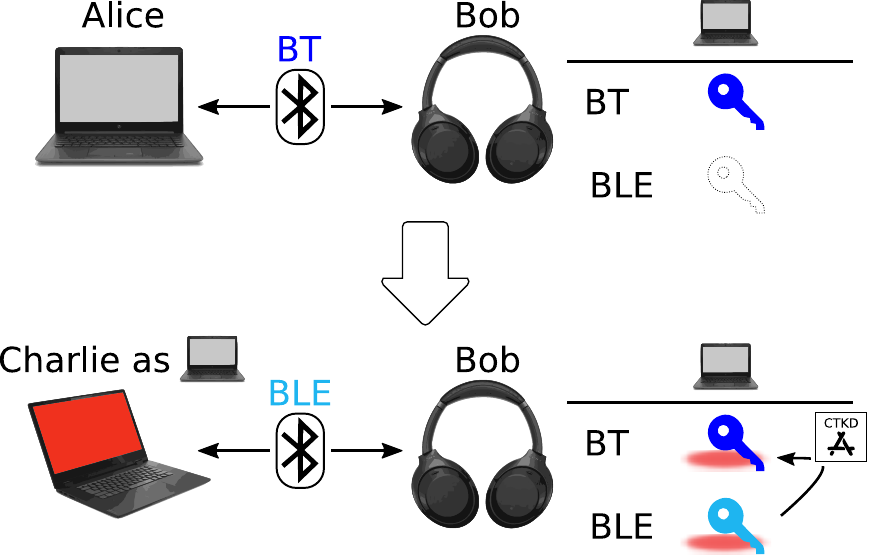}
    \caption{Attack strategy. Alice and Bob
    are paired over BT and run a secure BT session.
    Charlie pairs with Bob as Alice over BLE declaring CTKD support. Then
    Charlie agrees upon a BLE pairing key with Bob, and, via CTKD, tricks Bob
    into overwriting Alice's BT pairing key. As a result, Charlie can
    establish BT and BLE sessions with Bob as Alice, and
    takes over the real Alice who can no longer connect to Bob.
    Using a similar
    strategy, Charlie can also impersonate Bob to Alice, man-in-the-middle
    Alice and Bob, and establish unintended BT and BLE sessions as an
    arbitrary device.}
    \label{fig:imp-attacks}
\end{figure}

We now describe our attack strategy
with the help of Figure~\ref{fig:imp-attacks}. Let us assume that
Alice is a laptop and Bob is a pair of headphones and the victims are already
paired and they are running a secure BT session. Since the victims support
CTKD, they are also pairable over BLE, even if the transport is not currently
in use. Charlie sends a BLE pairing request to Bob (the victim) pretending to
be Alice, and claiming CTKD support. The attacker also declares \emph{no input/output
capabilities} to negotiated unauthenticated Just Works (JW) association. This
last step does \emph{not} trigger the key overwrite countermeasure described in
Section~\ref{sec:ctkd} as the attacker is neither changing the MitM protection
flag nor the strength (\ie\ entropy) of the pairing key.

Bob, even if running a BT session with Alice, has to answer to Charlie with
a BLE pairing response as Charlie's message is compliant with the Bluetooth
standard. Then, Charlie (as Alice) and Bob agree on a BLE pairing key and, via
CTKD, generate a new BT pairing key that \emph{overwrites} Alice's key in
Bob's BT key store.
In doing so, Charlie, wins two prizes with one shot, as he takes over Alice's
BT and BLE sessions with Bob. In other words, Alice can no longer connect
to Bob as she does not know the BT and BLE pairing keys (overwritten by the
attacker). Furthermore, Charlie also overwrites other security
keys that are distributed during pairing, including CSRK (signature key)
and IRK (MAC randomization key). We note that
the overwrite trick is transparent to the end user as the standard does not
mandate to notify the user about CTKD, and works even if Alice and Bob are
sharing BT \emph{and} BLE pairing keys before the attack takes place.

Following a similar strategy, Charlie can impersonate Bob to Alice,
man-in-the-middle them, and create unintended sessions as an arbitrary device
with a victim. We note that our attack strategy is effective because the
Bluetooth standard does not enforce important security properties at the
boundary between BT and BLE and does not address all cross-transport threats
in its threat model (see Section~\ref{sec:ctkd-vulns} for more details). In
the remaining of this section, we describe the technical details of the four
BLUR attacks.



\subsection{Impersonation Attacks}
\label{sec:blur-imp}

\begin{figure}[tb]
    \centering
    \includegraphics[width=1.0\linewidth]{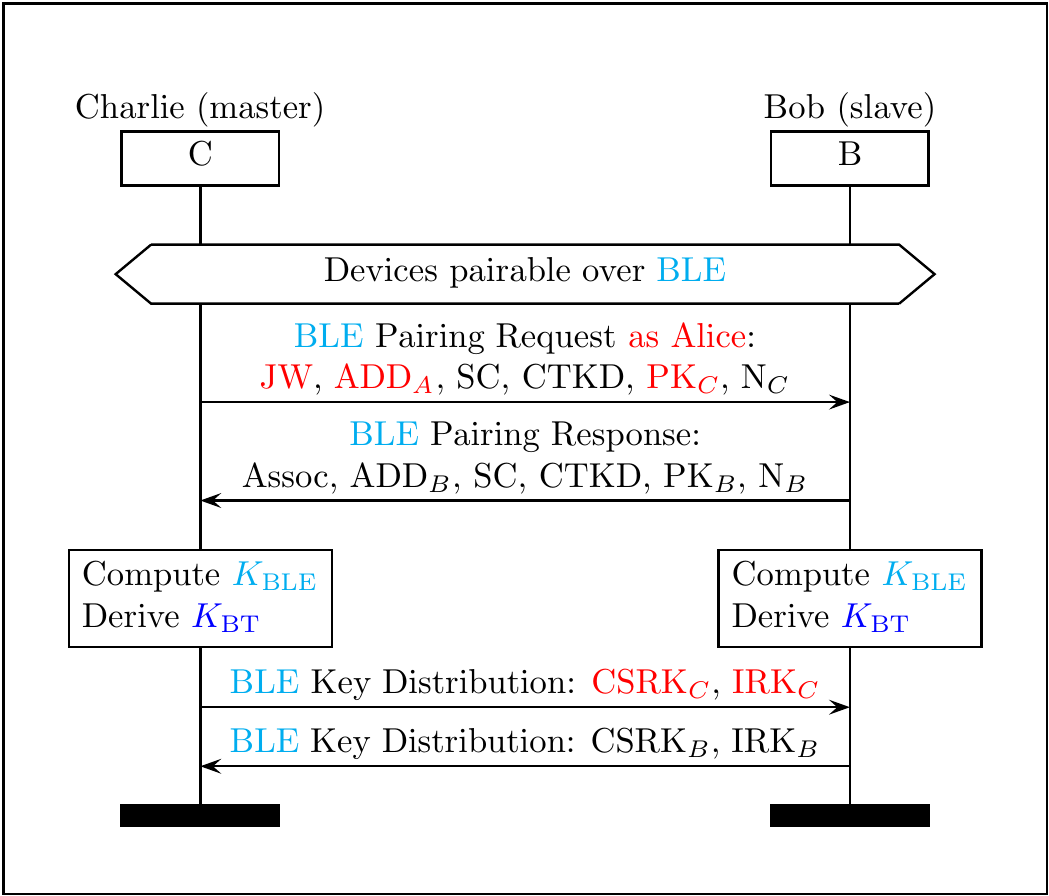}
    \caption{BLUR master impersonation attack. Charlie
    sends a BLE pairing request with Alice's address (ADD$_A$) including
    Just Works (JW) association, CTKD, and his public key
    (PK$_C$). Bob answers with a
    BLE pairing response thinking that he is talking to Alice.
    The attacker and the victim agree on \Kble, and
    derive \Kbt, via CTKD and complete BLE pairing by generating and
    distributing more keys over a secure BLE session. As a result of the
    master impersonation attack, Charlie
    tricks Bob into overwriting Alice's keys with his ones and takes over
    Alice who can no longer connect back to Bob.}
    \label{msc:ctkd-a2}
\end{figure}

\textbf{Master impersonation}
Charlie impersonates Alice and takes over her BT and BLE sessions with
Bob as in Figure~\ref{msc:ctkd-a2}. Bob is already paired with Alice, and
can run a BT session with her while Alice's impersonation takes place.
Notably, Bob must be pairable over BT and BLE to support CTKD from BT and BLE.
Charlie takes advantage of that and sends a BLE pairing request as Alice by using Alice's Bluetooth
address (ADD$_A$), Just Works (JW) association while
pairing, his public key (PK$_C$), and CTKD support.

As Charlie's BLE pairing request is standard-compliant,
Bob sends back a BLE pairing response believing that Alice wants to pair
(or re-pair) over BLE using CTKD. Then, Charlie and Bob compute \Kble,
derive \Kbt\ via CTKD, and exchange additional BLE key material
(\eg\ CSRK, IRK) over a BLE secure session. After the master impersonation
attack is completed Charlie takes over Alice's BT and BLE sessions by tricking Bob into
overwriting Alice's BT and BLE keys with his ones.

\begin{figure}[tb]
    \centering
    \includegraphics[width=1.0\linewidth]{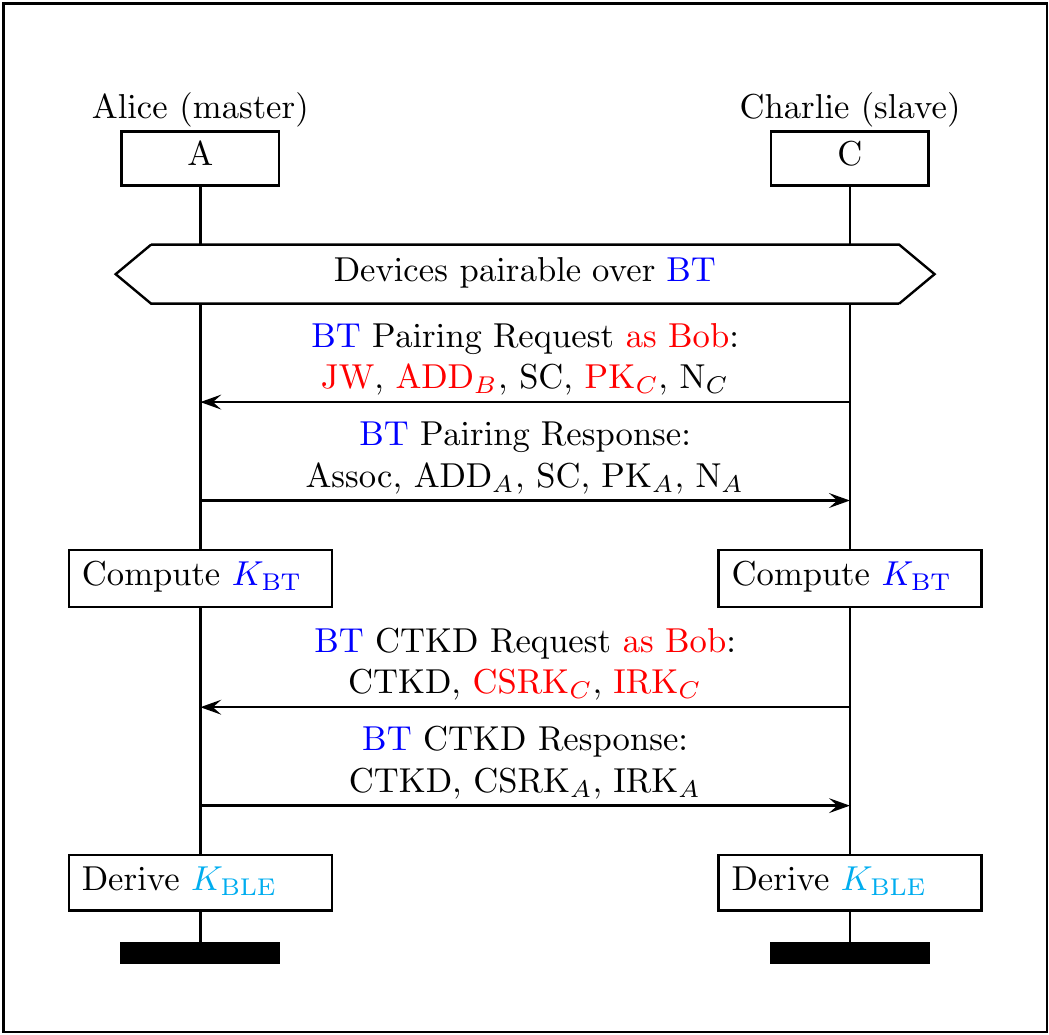}
    \caption{BLUR slave impersonation attack. Charlie sends a BT pairing
    request with Bob's address (ADD$_B$) including Just Works (JW)
    association, and his public key (PK$_C$). The
    pairing request is valid as BT enables to dynamically switch from slave to
    master before sending a pairing request. Alice answers with a BT pairing
    response believing that she is talking to Bob. The attacker and the victim
    establish \Kbt, negotiate CTKD and exchange additional keying material
    for BLE with a BT CTKD request and response messages, and
    derive \Kble. As a result of the slave impersonation attack, Charlie
    tricks Alice into overwriting Bob's keys with his ones and takes over
    Bob who can no longer connect back to Alice.}
    \label{msc:ctkd-a3}
\end{figure}

\textbf{Slave impersonation} Charlie impersonates Bob and takes
over his BT and BLE sessions with Alice as in Figure~\ref{msc:ctkd-a3}. Alice
and Bob have already paired and can run a BLE secure session while the
impersonation takes place. Alice has to be pairable over BT and BLE to provide
CTKD support from both transports, and Charlie takes
advantage of that by sending a BT pairing request to Alice as Bob
using Bob's address (ADD$_B$), Just Works (JW), and his
public key (PK$_C$). Charlie's pairing request is still standard-compliant even
if Charlie is supposed to be the slave as BT, unlike BLE, enables a slave to
switch to a master role before sending a pairing request.

Alice answers with a BT pairing response believing that
Bob wants to re-pair over BT, and the two agree on \Kbt. Then, Charlie starts a
secure BT session and sends a tunneled BLE pairing request to Alice still
pretending to be Bob. The BLE pairing request includes CTKD support and Charlie's
signature and MAC randomization BLE keys (CSRK$_C$, IRK$_C$).
Alice answers with a BLE pairing response tunneled over BT and the two derives
\Kble\ via CTKD.
Once the slave impersonation attack is completed, Charlie
takes over Bob's BT and BLE sessions by tricking Alice into overwriting Bob's
BT and BLE keys with his ones.

\begin{figure}[tb]
    \centering
    \includegraphics[width=.85\linewidth]{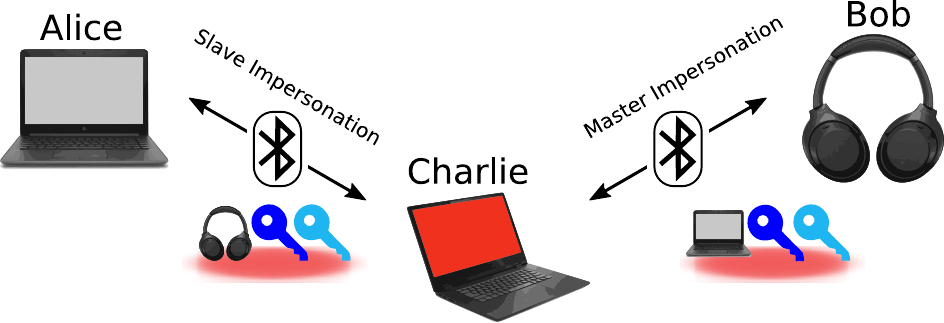}
    \caption{BLUR MitM attack. Charlie combines the master and
    slave impersonation attacks presented so far to establish a
    man-in-the-middle position between Alice and Bob both on BT and BLE.}
    \label{fig:mitm-attack}
\end{figure}

\textbf{Man-in-the-middle}
Charlie can conveniently combine the described master and slave attacks
to launch a cross-transport man-in-the-middle attack as shown in
Figure~\ref{fig:mitm-attack}. If Alice and Bob are running a BLE session,
Charlie starts with the slave impersonation attack presenting to Alice as Bob over
BT. Otherwise, he launches a master impersonation attack by targeting Bob as
Alice over BLE. After the first impersonation attack, the impersonated victim is
taken over and disconnects from the other victim. Then, Charlie targets the
impersonated victim with a second impersonation attack and establishes
a MitM position between the two victims. As a result, Charlie controls all
BT and BLE secure sessions between Alice and Bob.





\subsection{Unintended Session Attacks}
\label{sec:blur-anon}

CTKD enables unintended session attacks as it provides \emph{more ways} to
(silently) pair devices. In particular, two devices supporting CTKD are always
pairable over BT and BLE, but typically they use one transport at a
time. Hence, the attacker can target the unused transport as a stepping stone
to establish trusted bonds on BT and BLE via CTKD, while impersonating a random
device. To the best of our knowledge, this attack technique is new in the
context of Bluetooth.

\begin{figure}[tb]
    \centering
    \includegraphics[width=0.9\linewidth]{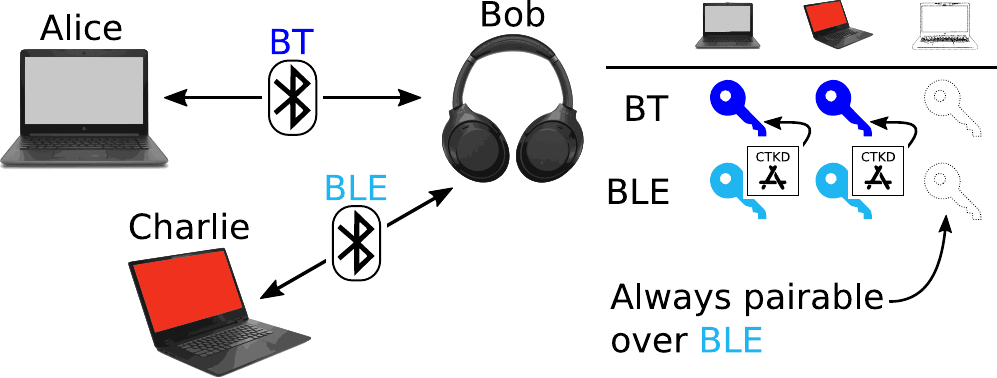}
    \caption{BLUR unintended sessions attack. Charlie can take advantage of
    CTKD to establish unintended BT and BLE session with Bob as a random
    device with arbitrary capabilities. The same can happen if Charlie targets
    Alice.}
    \label{fig:uns-attack}
\end{figure}

Let us see how an unintended session attack works in a scenario where
Alice and Bob are already paired and are running a secure BT session (see
Figure~\ref{fig:uns-attack}).
Charlie targets Bob by sending a BLE pairing request using a
random Bluetooth address, CTKD support, and Just Works for association. Bob
answers to Charlie's request and the two negotiate \Kble, and derive \Kbt\ via
CTKD. Now, Charlie can establish
secure but unintended BT and BLE sessions with Bob without breaking Bob's
existing sessions (\eg\ with Alice) and by using an anonymous
identity and arbitrary capabilities. Using a
similar strategy, Charlie can reach the same goals targeting Alice.

An unintended session attack is valuable for various reasons. It is
\emph{stealthy} as the attacker can establish a trusted relation with the
victim as an anonymous device and with minimal user interaction (\eg\ Just
Works). Moreover, it allows \emph{complete device enumeration} as the
attacker, being a trusted peer, can access all BT and BLE services, including
the protected ones unlike other attacks such as~\cite{celosia2019fingerprinting}.
Additionally, the attack deterministically
\emph{de-anonymizes} BLE devices, as the
attacker get access to the identity resolving key distributed during
pairing. Finally, it enables the attacker to reach
\emph{more} Bluetooth stack code sections than an untrusted device, including remote
code execution bugs in the pairing and secure session code.


\subsection{Attacks Discussion}
\label{sec:blur-discussion}

We now discuss the attacks' root causes, their effectiveness regardless of the
Bluetooth version, association method, and even CTKD support. Finally, we
detail how we discovered them.

\textbf{Root causes (CTIs)}
The BLUR attacks take
advantage of the four \emph{CTI} presented in Section~\ref{sec:ctkd-vulns}.
As shown in Table~\ref{tab:attack-issue},
a checkmark (\checkmark) indicates that a CTI is required,
an ''\emph{x}'' if it is not
needed, and an ''\emph{*}'' if it is sometimes needed.
From the table we see that all
attacks exploit extended pairability (CTI 1). The slave impersonation and
MitM attacks take advantage of role asymmetries (CTI 2), while some unintended
session attacks take advantage of that. Key tampering (CTI 3) is exploited
in all attacks as the attacker has to either write or overwrite keys using
CTKD. Association manipulation (CTI 4) is required in the first three attacks
when the victim expects a strong association mechanism but the attacker
negotiates Just Works.

\begin{table}[b]
  \renewcommand{\arraystretch}{1.3}
    \centering\small
    \begin{tabular}{@{}lccccc@{}}
        \toprule
                                    & CTI 1         & CTI 2       & CTI 3     & CTI 4      \\
        \midrule
        Master Impersonation        & \checkmark    & x          & \checkmark & * \\
        Slave Impersonation         & \checkmark    & \checkmark & \checkmark & *  \\
        MitM                        & \checkmark    & \checkmark & \checkmark & * \\
        Unintended Session          & \checkmark    & *          & \checkmark & x \\
        \bottomrule
    \end{tabular}
    \caption{Mapping the BLUR attacks to the CTI from
    Section~\ref{sec:ctkd-vulns}. CTI 1: extended pairing, CTI 2: role
    asymmetry, CTI 3: key tampering, and CTI 4:  association manipulation.
    We use a \checkmark\ if a CTI is required to
    conduct an attack, a x if is not required and a * if is only required in
    specific cases.}
    \label{tab:attack-issue}
\end{table}

\textbf{Bluetooth v5.1/5.2}
In Section~\ref{sec:ctkd} we describe a version-specific key overwrite
security argument included in the Bluetooth standard since v5.1. The Bluetooth
SIG currently uses this argument to state that the BLUR attacks are
\emph{not} effective against Bluetooth 5.1 and 5.2 devices (see
\url{https://tinyurl.com/vxhwftc2}).
We \emph{disagree} with
this statement, and we provided them empirical evidence
(see Section~\ref{sec:eval-results}) and solid arguments (described below)
about why this is not the case.

In particular, the key overwrite countermeasure is ineffective as it is
\emph{out of scope} with our attacks. Firstly, it does not cover \emph{non
key overwrite} cross-transport attacks such as the cross-transport
unintended session and key writing attacks presented in this work.
In addition, it does not protect against key overwrite
attacks \emph{not} involving the downgrade of keys' strength and MitM
protection. Specifically, our key overwrite attacks
declare "no input/output capabilities" to force the usage of Just Works
without downgrading key strengths or MitM protections.

\textbf{Association} The BLUR attacks are effective
regardless of the association methods supported by a victim, as the
attacker can \emph{always} downgrade it to Just Works. Even Just Works might
require minimal user interaction (\eg\ Yes/No pairing prompt) but remains an
\emph{unauthenticated} and vulnerable mechanism (\eg\ the user has no way to
tell if the remote pairing device is trustworthy).

\textbf{CTKD support}
Interestingly our attacks can be launched even if one of the victims does
\emph{not} support CTKD. By design CTKD's negotiation is not protected and enforced
across pairing sessions. Hence, if the attacker impersonates a device
not supporting CTKD, she can still use the BLUR attacks if the victim
supports CTKD. For example, the adversary can impersonate BT-only speakers
to a BT/BLE (dual-mode) laptop and exploit CTKD.

\textbf{Discovery}
We discovered the BLUR attacks by \emph{inference} from the public information
and RE details presented in Section~\ref{sec:ctkd-eval}. Our experiments
involved actual devices and static and dynamic analysis of the exchanged
Bluetooth packets and the CTKD code.




\section{Implementation}
\label{sec:impl}

In this section we describe our attack scenario, our implementation of a
custom attack device to perform the BLUR attacks and our re-implementation of
CTKD's key derivation function. We will open-source both implementations.

\subsection{Attack Scenario}
\label{sec:impl-as}

\begin{figure}[tb]
    \centering
    \includegraphics[width=0.9\linewidth]{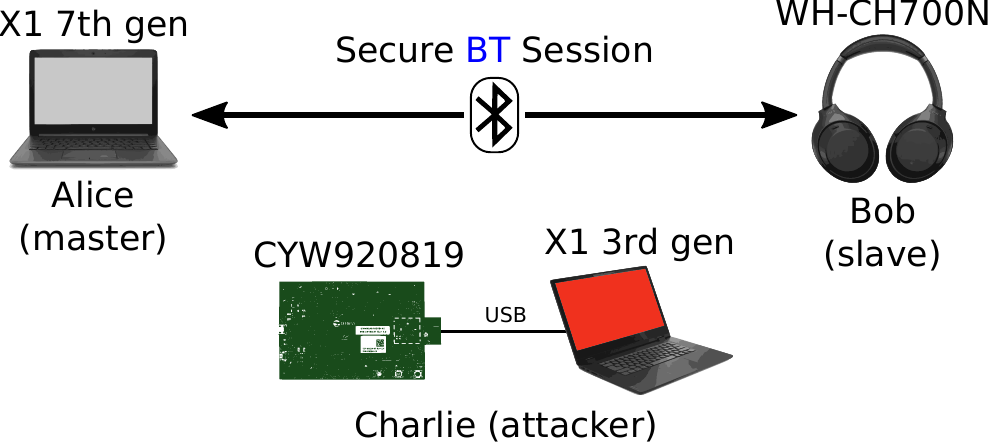}
    \caption{BLUR attack scenario. Alice (master) is a ThinkPad X1 7th gen,
    Bob (slave) is a pair of Sony WH-CH700N headphones and Charlie (attacker)
    is a CYW920819 board connected via USB to a ThinkPad X1 3rd gen. Alice and
    Bob have paired in absence of Charlie, and are running a secure BT session.}
    \label{fig:attack-scenario}
\end{figure}

Our attack scenario follows the example in Figure~\ref{fig:attack-scenario}
and includes two victims, Alice (master) and Bob (slave).
Alice is represented by a 7th
generation ThinkPad X1 laptop and Bob by a pair of Sony WH-CH700N
headphones. The attacker (Charlie) uses a CYW920819 development
board~\cite{cyw920819evb-02} and a 3rd generation ThinkPad X1 laptop as an
attack device. The implementation of the attack device is presented in
Section~\ref{sec:impl-ad}. In our evaluation, presented in
Section~\ref{sec:eval}, we use the same
attack scenario to attack other victim devices.

Table~\ref{tab:as} summarizes the most relevant features of Alice, Bob, and
Charlie. Alice and Bob have an Intel Bluetooth chip, while Bob has a Cambridge
Silicon Radio (CSR) one. Alice, Bob, and Charlie support respectively
Bluetooth 5.1, 4.1, and 5.0. Alice and Charlie support Secure Connections both
on the Host and the Controller, while Bob only on the Controller. All devices
support BT, BLE, and CTKD. Regarding pairing association methods, the
laptops support Numeric Comparison, while the headsets
only support Just Works as they lack a display.

\begin{table}[tb]
  \renewcommand{\arraystretch}{1.3}
    \centering\small
    \begin{tabular}{@{}llll@{}}
        \toprule
                      & Alice         & Bob               & Charlie\\
        \midrule
        Device(s)     & X1 7th gen    & WH-CH700N         & \parbox[c]{1.7cm}{X1 3rd gen / CYW920819} \\
        Radio Chip    & Intel         & CSR               & Intel / Cypress  \\
        Subversion    & 256           & 12942             & 256 / 8716  \\
        Version       & 5.1           & 4.1               & 5.0  \\
        Name          & x7            & WH-CH700N         & x1  \\
        ADD         & Redacted      & Redacted          & Redacted  \\
        Class         & 0x1c010c      & 0x0               & 0x0  \\
        BT SC         & True          & Only Controller   & True  \\
        BT AuthReq    & 0x03          & 0x02              & 0x03  \\
        BLE SC        & True          & True              & True  \\
        BLE AuthReq   & 0x2d          & 0x09              & 0x2d\\
        CTKD          & True          & True              & True  \\
        h7            & True          & False             & True  \\
        Role          & Master        & Slave             & Master  \\
        IO            & Display       & No IO             & Display  \\
        Association   & Numeric C.    & Just Works        & Numeric C.  \\
        Pairable      & True          & True              & True  \\
        \bottomrule
    \end{tabular}
    \caption{Relevant features of Alice, Bob, and Charlie.
    We redact the devices' Bluetooth addresses for privacy reasons.}
    \label{tab:as}
\end{table}


\subsection{Attack Device}
\label{sec:impl-ad}

\begin{figure}[tb]
    \centering
    \includegraphics[width=0.9\linewidth]{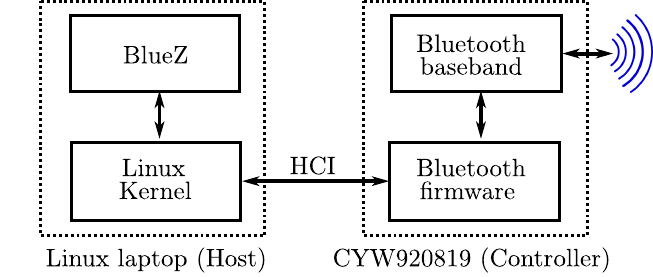}
    \caption{Attack device block diagram. The attack device is composed of
    Linux laptop (Host) and a CYW920819 development board (Controller)
    connected via USB and
    communicating using the Host Controller Interface (HCI) protocol.}
    \label{fig:attack-device}
\end{figure}

To conduct our attacks we developed an attack device making use
of a \emph{CYW920819 development board} connected to a \emph{Linux laptop} (see
Figure~\ref{fig:attack-device}). The devices support BT, BLE, SC, and
CTKD. We picked these devices as COTS devices do not allow
to modify their Bluetooth firmware (Controller) but at most the OS Bluetooth
stack (Host).
A software-defined radio (SDR) is also out of scope because there is no
open-source BT/BLE SDR stack currently available.

Instead, with our attack device, we can program our development board
(Bluetooth Controller) to impersonate any BT/BLE device, we can patch its
closed-source firmware to control both BT LMP and BLE LL link layer packets.
Moreover, we
can alter the laptop's BT and BLE kernel and user-space code to set Bluetooth
Host-specific configuration bits such as negotiating CKTD and Just Works. We
now describe in detail how we modify the attack device's Host and Controller
components.

\textbf{Host modifications}
For the host, we use standard Linux tools to configure
an Bluetooth interface (\eg\ \texttt{hciconfig}), and to discover and pair with a device
(\eg\ \texttt{bluetoothctl}, \texttt{hcitool} and \texttt{btmgmt}). In particular,
\texttt{btmgmt} was very useful as it provides handy low-level commands. For
example, it includes commands to toggle BT, BLE, SC, scanning, and
advertising. Moreover, it allows to easily send custom pairing requests on BT
and BLE and to set the related association (\eg\ Just Works).

Furthermore, we configured our host to get all link-layer packets
sent and received by the controller. This is handy as it enables to monitor both
HCI and link-layer packets directly from the host (\eg\ using Wireshark).
To activate link-layer packet forwarding, we
sent a proprietary Cypress HCI command from the host to the controller that switches
on an undocumented diagnostic mode in the controller. Then, we added extra C code to
the Linux kernel to parse those special HCI packets in the host.

\textbf{Controller modifications} We modified the controller by dynamically
patching the development board Bluetooth firmware using a Cypress proprietary
mechanisms. To patch the firmware we had to extract it from the board and
statically reverse-engineer its relevant parts. In particular,
to extract the firmware we used a
proprietary HCI command to read and save a runtime RAM snapshot from the board's SoC.
We use the memory maps that we extracted from the board's SDK to extract the
memory segments from the snapshot (\eg\ ROM, RAM, and the scratchpad).
As expected, the firmware was in the ROM segment and was a stripped ARM binary
containing 16-bit Thumb instructions.

To reverse-engineer the firmware, we loaded
the ROM, RAM, and scratchpad in Ghidra and statically analyzed them. In our first
pass, we isolated the libc functions (\eg\ \texttt{malloc} and
\texttt{calloc}) by looking at the signatures and the code patterns of the functions
that are called the most. Then, we found the firmware debugging
symbols hidden in the board's SDK and loaded them into Ghidra. Using these
symbols we isolated functions and data structures relevant to the BLUR
attacks. Then, we wrote ARM Thumb assembly patches to change their behaviors and
we apply those patches at runtime using internalblue~\cite{internalblue}, an
open-source toolkit to manage several Bluetooth devices including our board. Our
set of patches allows transforming our board in whatever device we want by
changing its identifiers including addresses, names, and capabilities,


\subsection{CTKD Key Derivation Function (KDF)}
\label{sec:impl-ctkd}

We implemented CTKD's custom KDF, following the Bluetooth
standard~\cite[p.~1401]{btsig-52}. This implementation is \emph{not} required
to conduct the attack, but it was used to check that the CTKD keys were
correctly derived. Our implementation is written in Python 3, uses the PyCA
cryptographic module~\cite{pyca-crypto}, and was successfully tested against
the test vectors in the standard~\cite[p.~1721]{btsig-52}. We now describe its
technical details.

The KDF computes \Kble\ using \Kbt, ``tmp2'' and ``brle'',
and \Kbt\ from \Kble, ``tmp1'' and ``lebr''. Each case can be represented by a
system of equations (see below). If CTKD is run from
BT then the first system of equations is used
otherwise the second. Each system has two equations
and the top equation is
used if both devices support the \emph{h7} key conversion function. h7
is negotiated during pairing using the AuthReq flag~\cite[p.~1634]{btsig-52}.
All four equations internally use \emph{f$(a,b)$} that is implemented as
AES-CMAC$(key, plaintext)$.

\[ K_{BLE} =
\begin{cases}
    f\left(f\left(tmp2, K_{BT}\right), brle\right)       & \quad \text{if h7 is supported}\\
    f\left(f\left(K_{BT}, tmp2\right), brle\right)       & \quad \text{otherwise}
\end{cases}
\]

\[
K_{BT} =
\begin{cases}
    f\left(f\left(tmp1, K_{BLE}\right), lebr\right)       & \quad \text{if h7 is supported}\\
    f\left(f\left(K_{BLE}, tmp1\right), lebr\right)       & \quad \text{otherwise}
\end{cases}
\]





\section{Evaluation}
\label{sec:eval}

In this section we present how we successfully conducted the BLUR attacks
on \devices\ devices using \chips\ unique Bluetooth chips.
Our results confirm that the BLUR attacks are effective against different
device types
(\eg\ laptops, smartphones, headphones, and development boards), manufacturers
(\eg\ Samsung, Dell, Google, Lenovo, and Sony), operating systems (\eg\
Android, Windows, Linux, and proprietary OSes), and Bluetooth firmware (\eg\
Broadcom, CSR, Cypress, Intel, Qualcomm, and Samsung).

\begin{table*}[tb]
  \renewcommand{\arraystretch}{1.3}
  \newlength{\circwidth}
  \settowidth{\circwidth}{\cone}
    \centering\small
    \begin{tabular}{@{}lllllclccc@{}}
        \toprule
        \multicolumn{3}{@{}c@{}}{Device} & \multicolumn{2}{@{}c@{}}{Chip} &    Bluetooth        & \multicolumn{4}{@{}c@{}}{BLUR Attack} \\
        \cmidrule(r){1-3} \cmidrule(r){4-5} \cmidrule(r){6-6} \cmidrule{7-10}
        Producer & Model        & OS          & Producer& Model        & Version & Role   & \makebox[\circwidth]{MI/SI} & MitM    & US \\
        \midrule
        Cypress & CYW920819EVB-02 & Proprietary & Cypress  & CYW20819        & 5.0     & Slave  & \checkmark  & \checkmark & \checkmark\\
        Dell    & Latitude 7390 & Win 10 PRO  & Intel      & 8265            & 4.2     & Slave  & \checkmark  & \checkmark & \checkmark\\
        Google  & Pixel 2       & Android     & Qualcomm   & SDM835          & 5.0     & Slave  & \checkmark  & \checkmark & \checkmark\\
        Google  & Pixel 4       & Android     & Qualcomm   & 702             & 5.0     & Slave  & \checkmark  & \checkmark & \checkmark\\
        Lenovo  & X1 (3rd gen)  & Linux       & Intel      & 7265            & 4.2     & Slave  & \checkmark  & \checkmark & \checkmark\\
        Lenovo  & X1 (7th gen)  & Linux       & Intel      & 9560            & 5.1     & Slave  & \checkmark  & \checkmark & \checkmark\\
        Samsung & Galaxy A40    & Android     & Samsung    & Exynos 7904     & 5.0     & Slave  & \checkmark  & \checkmark & \checkmark\\
        Samsung & Galaxy A51    & Android     & Samsung    & Exynos 9611     & 5.0     & Slave  & \checkmark  & \checkmark & \checkmark\\
        Samsung & Galaxy A90    & Android     & Qualcomm   & SDM855          & 5.0     & Slave  & \checkmark  & \checkmark & \checkmark\\
        Samsung & Galaxy S10    & Android     & Broadcom   & BCM4375         & 5.0     & Slave  & \checkmark  & \checkmark & \checkmark\\
        Samsung & Galaxy S10e   & Android     & Broadcom   & BCM4375         & 5.0     & Slave  & \checkmark  & \checkmark & \checkmark\\
        Samsung & Galaxy S20    & Android     & Broadcom   & BCM4375         & 5.0     & Slave  & \checkmark  & \checkmark & \checkmark\\
        Xiaomi  & Mi 10T Lite   & Android     & Qualcomm   & 9312            & 5.1     & Slave  & \checkmark  & \checkmark & \checkmark\\
        Xiaomi  & Mi 11         & Android     & Qualcomm   & 10765           & 5.2     & Slave  & \checkmark  & \checkmark & \checkmark\\
        Sony    & WH-1000XM3    & Proprietary & CSR        & 12414           & 4.2     & Master & \checkmark  & \checkmark & \checkmark\\
        Sony    & WH-CH700N     & Proprietary & CSR        & 12942           & 4.1\makebox[0pt][l]{$^\dagger$}     & Master & \checkmark  & \checkmark & \checkmark\\
        \bottomrule
      \multicolumn{10}{@{}l@{}}{\scriptsize $\dagger$ CTKD was backported by the vendor to Bluetooth 4.1.}
    \end{tabular}
    \caption{BLUR attacks evaluation results. The first three columns show the
    device's producer, model, and OS. The next two columns state the Bluetooth
    chip's producer and model. The sixth column tells the Bluetooth version of
    the target device. The seventh column indicates the attacker role.
    The last three columns contain a checkmark (\checkmark)
    if a device is vulnerable to the relevant BLUR attack.}
    \label{tab:results}
\end{table*}

\subsection{Setup}
\label{sec:impl-attacks}

The BLUR attacks, presented in Section~\ref{sec:blur}, include master
impersonation, slave impersonation, man-in-the-middle, and unintended
session attacks. In the next paragraphs, we describe how we conducted
each attack using the attack device described in Section~\ref{sec:impl-ad}.

\textbf{Laptop (master) BLUR impersonation attack} To impersonate the laptop, we
patch our attack device to clone the laptop's  Bluetooth features
(\eg\ Bluetooth address, name, device class, and security parameters)
Then, we send a BLE
pairing request from the attack device to the headphones declaring CTKD and
Just Works support. The malicious BLE pairing request is sent
using \texttt{btmgmt}'s text-based
user interface (TUI). The headphones accept the pairing request, and the
devices agree on \Kble, derive \Kbt\ via CTKD and
establish a secure BLE session. Then, the headphones
terminate the BT session with the impersonated laptop and establish a secure
BT session with the attack device. The impersonated laptop cannot connect back
with the headphones as it
does not possess the correct pairing keys overwritten by the attacker.

\textbf{Headphones (slave) BLUR impersonation attack} To impersonate the
headphones, we patch our attack device to clone the headphones' Bluetooth
features. Then, we send a BT pairing request from the attack
device to the laptop declaring CTKD and Just Works support using \texttt{btmgmt}'s TUI.
The laptop accepts to pair over BT as a BLE slave can send a BT pairing
request as a master. The devices agree on \Kbt, derive \Kble via CTKD,
and establish a secure session over BT. The impersonated headphones cannot connect to the
laptop as they do not own the correct pairing keys.



\textbf{BLUR Man-in-the-middle attack} By using
two development boards connected to the same laptop, we can impersonate the laptop and the
headphones at the same time, and man-in-the-middle them.
In particular, we run the laptop (master) impersonation attack first, and then
the headphone (slave) impersonation attack. As a result, the attack device
positions itself in the middle between the victims.

\textbf{BLUR Unintended sessions attack} For the unintended session attack,
we patched our attack device to look like an unknown device to the current
victim (\eg\ unknown Bluetooth address and name). If the victim is a master,
we run the same steps used in the slave
impersonation attack otherwise we use the master impersonation attack's steps.
In both cases, the attacker completes pairing using CTKD and can establish
secure sessions over BT and BLE with the victim.


\subsection{Results}
\label{sec:eval-results}

We exploited the BLUR attacks against \devices\ unique devices employing
\chips\ different Bluetooth chips and covering all Bluetooth versions
supported by CTKD. We show our results in
Table~\ref{tab:results}. The table's first six columns indicate the device's
producer, model name, operating system, chip manufacturer, chip model, and
Bluetooth version. The seventh column contains either Slave if the attacker's
role is slave, or Master otherwise. The table's last three columns
have a checkmark (\checkmark) if a device is vulnerable to master or slave
impersonation attack (MI/SI), MitM, or unintended session (US) attack.

Table~\ref{tab:results} empirically demonstrates that the BLUR attacks are
\emph{effective} on actual devices and are \emph{compliant} with the Bluetooth
standard.
The attacks are effective on all Bluetooth versions
supporting CTKD (\ie\ Bluetooth 4.2, 5.0, 5.1, and 5.2). In addition, they
succeed regardless of the hardware and software details of the victim device.
Interestingly, they work even on older Bluetooth versions to which CTKD was
backported.

Moreover, Table~\ref{tab:results} experimentally confirms that Bluetooth
5.1/5.2 are \emph{still vulnerable} to the BLUR attacks, despite the CTKD key
overwrite mitigation in the standard~\cite[p.~1401]{btsig-52}. See
Section~\ref{sec:ctkd} for an introduction and
Section~\ref{sec:blur-discussion} for our explanation about why it is
ineffective. We want to evaluate more 5.1/5.2
devices supporting CTKD, such as speakers and headsets. So far,
we were able to find only 5.1/5.2 high-end devices supporting CTKD
(\eg\ Xiaomi Mi 11, Mi 10T Lite, and ThinkPad X1 7th gen).



\section{Discussion}
\label{sec:discussion}


\subsection{Countermeasures}
\label{sec:counter}

To \emph{concretely} address the BLUR attacks and their root causes
(CTIs), we present four countermeasures. As they act at \emph{the protocol level},
they are effective regardless of the Bluetooth version number, and as they
also cover non key-downgrade attacks, they block all BLUR attacks. This is not
the case with the Bluetooth SIG's 5.1 key overwrite countermeasure described in
Section~\ref{sec:ctkd}.

\textbf{C1: Disable pairing when not needed}
To prevent an attacker from pairing with a device on unused transports, a
device should automatically stop being pairable on a transport that is not
currently in use. To avoid DoS issues, a device should also allow a user to
enable and disable pairing manually on a specific transport.

\textbf{C2: Align BT and BLE roles}
To fix role asymmetries when using CTKD from BT or BLE, a device should
store the transport and the role used while pairing, and enforce it across
re-pairings. In case of a role mismatch,
the device should abort pairing. We note that the BIAS
paper~\cite{antonioli20bias} also takes advantage of role switching but is not
proposing role switch enforcement as a countermeasure.


\textbf{C3: Prevent cross-transport key tampering}
To prevent cross-transport key overwrites via CTKD, a device should disable
it if a pairing key already exists for the other transport. To overwrite a
pairing key a user should explicitly re-pair on that transport. To mitigate
cross-transport key writes, CTKD should be disabled when two devices, who
already share a pairing key on a transport, re-pair on that transport with
a weaker key (that would be used as input to CTKD). A key is stronger than
another one if its entropy is higher or if is established with a stronger
association mechanism.

\textbf{C4: Enforce strong association mechanisms}
To prevent an attacker from manipulating associations across transports,
a device should keep track of the association
mechanism used while pairing for the first time and enforce
it for subsequent re-pairings (across BT and BLE). There is no reason why
two devices which support strong association would want to ever use a weaker
association scheme. If a weaker mechanism than the one stored is proposed,
pairing should be aborted.

The four countermeasures ultimately block the BLUR attacks. In
particular, C3 prevents impersonation and MitM as the attacker cannot
write or overwrite keys across transports, but only target
BT and BLE separately. To stop the unintended session attacks,
C1 is also needed as the
attacker should not be able to pair with CTKD on unused transports. C2 and C4
help mitigate the attacks by providing more defense-in-depth, but
they are not strictly required.

Our countermeasures are easy to implement and do not rely on
backward-incompatible features. In particular, they
can be implemented in the \emph{Bluetooth Host} (\ie\
OS level). C2, C3, and C4 can be realized by keeping track of extra data
that is exchanged during pairing (\eg\ device role,
association) and aborting the protocol when needed. Logging is
already supported and used by the Host (\eg\ to store pairing keys).
C1 can be implemented with
a timer that disables pairability on a transport when not needed and a simple
user interface to monitor and switch on/off pairability for BT and BLE.

\textbf{PoC for C3}
To verify the effectiveness of C3 and show that our mitigations are easy to
implement and do not impact normal operations, we developed a proof-of-concept (PoC) for C3 for Linux.
We can evaluate multiple device classes simultaneously by testing Linux, as
it is employed by Android smartphones, embedded devices, and laptops.
The C3 PoC works as follows. We pair a Linux laptop (victim) with an arbitrary
device with CTKD. Then, to disable key overwrites on the laptop, we unset the 
write permission bit of the pairing key file stored at \texttt{/var/lib/bluetooth}.
We then use the paired devices normally to demonstrate no impact on benign use.
Finally, we run the BLUR impersonation attack and we confirm that it is
ineffective as the attacker cannot overwrite the pairing keys.


\subsection{Lessons Learned}
\label{sec:ll}


\textbf{Specification and modeling}
Security mechanisms that \emph{cross} the security boundary
between two technologies should be well-specified and tested against
a comprehensive cross-transport threat model. On the contrary, the Bluetooth
standard provides an incomplete specification for CTKD and
only discusses some cherry-picked cross-transport threats.

\textbf{Security guarantees}
Cross-transport mechanisms should
be designed such that the mechanisms trusted at the boundary between the two
transport (\ie\ BT and BLE pairing) have the \emph{same} threat model and provides
\emph{equivalent} security guarantees. This is not the case for Bluetooth as
BT and BLE use different pairing protocols, link layer mechanisms, and threat models.

\textbf{Usability vs. Security}
CTKD was introduced to improve Bluetooth's usability, but, in light of the
presented attacks, the usability benefits are not balancing the security
issues deriving from CTKD. Indeed, it is paramount to find a good balance
between usability and security and \emph{not} trade off the latter for the former.

\section{Related Work}
\label{sec:related}

\begin{table*}[tb]
  \centering\small
  \begin{tabular}{@{}lllllc@{}c@{}c@{}cccl@{}}
    \toprule
    &&& & \multicolumn{7}{c}{Attack}\\\cmidrule{5-11}
    & Year & Paper                                            & Target  & Phase   & C     & I     & A     & K     & SC/SCO     & Persistent     & Note\\
    \midrule
    \multicolumn{11}{@{}l}{\emph{Attacks on BT}}\\
    & 2016 & Albazrqaoe et al.~\cite{albazrqaoe2016practical} & \design & Any     & \Xpar & \Xno  & \Xno  & \Xno  & x          & \notpersistent & BlueEar Sniffer \\
    & 2017 & Seri et al.~\cite{armis2017blueborn}             & \impl   & Pairing & \Xyes & \Xyes & \Xyes & \Xno  & NA         & \persistent    & BlueBorne\\
    & 2018 & Sun et al.~\cite{sun2018man}                     & \design & Pairing & \Xyes & \Xyes & \Xyes & \Xno  & \checkmark & \notpersistent & Passkey (MitM)\\
    & 2018 & Biham et al.~\cite{biham2018breaking}            & \impl   & Pairing & \Xyes & \Xyes & \Xyes & \Xpar & NA         & \persistent    & Fixed Coordinate Invalid Curve\\
    & 2019 & Ossmann et al.~\cite{ubertooth}                  & \design & NA      & \Xpar & \Xno  & \Xno  & \Xno  & x          & \notpersistent & Ubertooth sniffer\\
    & 2019 & Antonioli et al.~\cite{antonioli19knob}          & \design & Pairing & \Xyes & \Xyes & \Xpar & \Xno  & \checkmark & \notpersistent & KNOB (MitM) \\
    & 2020 & Antonioli et al.~\cite{antonioli20bias}          & \design & Pairing & \Xyes & \Xyes & \Xyes & \Xno  & \checkmark & \notpersistent & BIAS \\
    & 2021 & Tschirschnitz et al.~\cite{von21method}          & \design & Pairing & \Xyes & \Xyes & \Xyes & \Xno  & \checkmark & \notpersistent & Method Confusion (MitM)\\
    \multicolumn{11}{@{}l}{\emph{Attacks on BLE}}\\
    & 2016 & Jasek et al.~\cite{jasek2016gattacking}          & \design & NA      & \Xpar & \Xno  & \Xno  & \Xno  & x          & \notpersistent & Black Hat \\
    & 2019 & Seri et al.~\cite{armis2019bleedingbit}          & \impl   & NA      & \Xno  & \Xpar & \Xpar & \Xno  & NA         & \persistent    & Bleedingbit \\
    & 2020 & Zhang et al.~\cite{zhang2020breaking}            & \design & Pairing & \Xpar & \Xpar & \Xpar & \Xno  & \scon      & \notpersistent & MitM (SCO) \\
    & 2020 & Wu et al.~\cite{wu20blesa}                       & \design & Session & \Xno  & \Xno  & \Xyes & \Xno  & \checkmark & \notpersistent & BLESA \\
    & 2020 & Garbelini et al.~\cite{garbelini2020sweyntooth}  & \impl   & Any     & \Xpar & \Xpar & \Xpar & \Xno  & NA         & \notpersistent & SweynTooth fuzzer \\
    & 2020 & Antonioli et al.~\cite{antonioli20tops}          & \design & Pairing & \Xyes & \Xyes & \Xpar & \Xno  & \checkmark & \notpersistent & Downgrade (MitM) \\
    \multicolumn{11}{@{}l}{\emph{ Cross-transport attacks on BT and BLE}}\\
      & 2021 & \textbf{BLUR (this work)}                               & \design & Any
      & \Xyes & \Xyes & \Xyes & \Xpar & \scon      & \persistent    & \textbf{First against CTKD}\\
    \bottomrule
  \end{tabular}
  \caption{Comparison with related work. The BLUR attacks are the first
    \emph{cross-transport} standard-compliant attacks for Bluetooth and the first
    targeting \emph{CTKD}. C = Data Confidentiality, I = Data Integrity, A = Device Authentication, K = Key disclosure. No (\Xno) Partially (\Xpar), Yes (\Xyes).}
  \label{tab:related}
\end{table*}

Bluetooth provides a royalty-free and widely-available cable replacement
technology~\cite{haartsen1998bluetooth}. Standard-compliant attacks on
Bluetooth are particularly dangerous as all devices are affected, regardless
of version numbers or implementation details. Such attacks were discovered
since Bluetooth v1.0~\cite{jakobsson2001security,lindell2008attacks}.

The Bluetooth standard evolved over time to include better pairing mechanisms
(\eg\ SSP, SC) and two transports (\eg\ BT, BLE).
Recent standard-compliant attacks on BT include
attacks on legacy pairing~\cite{shaked2005cracking},
secure simple pairing
(SSP)~\cite{haataja2010two,sun2018man,biham2018breaking},
association~\cite{hypponen2007nino,von21method},
key negotiation~\cite{antonioli19knob},
and authentication procedures~\cite{levi2004relay,wong2005repairing,antonioli20bias}.
Regarding-BLE we have attacks on legacy
pairing~\cite{ryan2013bluetooth}, key negotiation~\cite{antonioli20tops},
SSP~\cite{biham2018breaking,zhang2020breaking},
reconnections~\cite{wu20blesa}, and GATT~\cite{jasek2016gattacking}.

The BLUR attacks are novel compared to \emph{prior} standard-compliant
attacks. As we can see from Table~\ref{tab:related} they are the first
\emph{cross-transport} attacks, meaning the first targeting BT from BLE and
vice versa. Moreover, no prior attacks targeted (and evaluated the security
of) CTKD. Finally, like the BIAS attack~\cite{antonioli20bias}, they require a
weak threat model as the attacker can target a victim at any time. Unlike the
BIAS attack, the effect of our attacks is persistent across sessions. Like
other standard-compliant attacks, the BLUR attacks are effective regardless of
the security mode (\eg\ SSP with SC), association method (\eg\ Numeric
Comparison), and Bluetooth version numbers.

We have seen attacks targeting specific implementation
flaws on BT~\cite{armis2017blueborn} and
BLE~\cite{armis2019bleedingbit,garbelini2020sweyntooth}. As our attacks
target the specification level, they are effective regardless of the
implementation details.
%
%
Several surveys on BT and BLE security were
published~\cite{dunning2010taming,minar2012bluetooth,padgette2017guide} but
neither of those surveys nor the Bluetooth standard considers CTKD as
a threat. We here demonstrate that CTKD is a serious threat and must be
included in the standard Bluetooth threat model.


Cross-protocols attacks were exploited for proximity technologies
using Bluetooth and Wi-FI. Two prominent examples are attacks
on Apple ZeroConf~\cite{bai2016staying} and Google Nearby
Connections~\cite{antonioli19rearby}. However, no prior attack
targeted the BT/BLE combination.


The cryptographic primitives used by Bluetooth have been extensively
analyzed. For example, the \Ezero\ cipher used by BT was
investigated~\cite{fluhrer2001analysis} and it is considered relatively
weak~\cite{padgette2017guide}. SAFER+, used for authentication, was analyzed
as well~\cite{kelsey1999key}. BT and BLE ``Secure Connections'' use the
AES-CCM authenticated-encryption cipher. AES-CCM was extensively analyzed
~\cite{jonsson2002security, rogaway2011evaluation} and it is FIPS-compliant.
As our attacks are at the protocol-level, they
are effective even with perfectly secure cryptographic primitives.






\section{Conclusion}
\label{sec:conclusion}

This work presents the first security evaluation of CTKD. CTKD was introduced
in the Bluetooth standard to improve the usability of pairing.
With CTKD two devices can pair on BT (or BLE) and generate pairing keys for both
transports. CTKD is a novel attack surface as it allows
to tamper with BT from BLE and vice versa, and is only partially documented in the Bluetooth standard (without an appropriate security analysis).

To address these issues, we RE the CTKD protocols and analyzed them using a
\emph{cross-transport} attacker model.
Our analysis uncovers four critical cross transport
issues (CTI) in the specification of CTKD. As such, \emph{all} Bluetooth
devices supporting CTKD are currently affected by those vulnerabilities.

We leverage the CTIs to implement four standard-compliant cross-transport attacks.
Our attacks allow an attacker to impersonate and MitM  devices, and allow establishing unintended (anonymous) sessions with a victim to enumerate sensitive data and send malicious packets. The attacks are the first standard-compliant BT and BLE attacks to not require the attacker to be present when a victim is pairing or establishing a secure session, unlike prior work~\cite{hypponen2007nino, haataja2010two, ryan2013bluetooth, sun2018man,
biham2018breaking, antonioli19knob, antonioli20tops,
antonioli20bias,wu20blesa,zhang2020breaking,von21method}. In particular, our
attacks are the first that can be conducted in absence of one of the victims.
The attacks are effective \emph{regardless of} the targeted Bluetooth version and security mode (\eg\ SSP, SC, on strong association). 

To demonstrate the practicality of the BLUR attacks, we presented a low-cost
implementation based on readily available hardware and open-source software.
We use our implementation to empirically confirm that the BLUR attacks
are standard-compliant and effective all targeted devices. In particular, we
exploited \devices\ different devices using \chips\ unique Bluetooth chips.
Our device sample includes all Bluetooth versions supporting CTKD (\eg\ 4.2,
5.0, 5.1, and 5.2) and BT and BLE devices supporting SC and strong association.

To fix the presented attacks and their root causes, we propose protocol-level
countermeasures, and demonstrate the efficacy of the most important one
(disable key overwrites) experimentally.



\bibliographystyle{plain}
\bibliography{bibliography}

\begin{thebibliography}{10}

\bibitem{albazrqaoe2016practical}
Wahhab Albazrqaoe, Jun Huang, and Guoliang Xing.
\newblock Practical bluetooth traffic sniffing: Systems and privacy
  implications.
\newblock In {\em Proceedings of the Annual International Conference on Mobile
  Systems, Applications, and Services}, pages 333--345. ACM, 2016.

\bibitem{antonioli19knob}
Daniele Antonioli, Nils~Ole Tippenhauer, and Kasper Rasmussen.
\newblock The {KNOB} is broken: Exploiting low entropy in the encryption key
  negotiation of {Bluetooth BR/EDR}.
\newblock In {\em Proceedings of the USENIX Security Symposium}. USENIX, August
  2019.

\bibitem{antonioli19rearby}
Daniele Antonioli, Nils~Ole Tippenhauer, and Kasper Rasmussen.
\newblock {Nearby Threats: Reversing, Analyzing, and Attacking Google's
  ``Nearby Connections'' on Android}.
\newblock In {\em Proceedings of the Network and Distributed System Security
  Symposium ({NDSS})}, February 2019.

\bibitem{antonioli20bias}
Daniele Antonioli, Nils~Ole Tippenhauer, and Kasper Rasmussen.
\newblock {BIAS: Bluetooth Impersonation AttackS}.
\newblock In {\em Proceedings of Symposium on Security and Privacy (S\&P)}.
  IEEE, May 2020.

\bibitem{antonioli20tops}
Daniele Antonioli, Nils~Ole Tippenhauer, and Kasper Rasmussen.
\newblock {Key Negotiation Downgrade Attacks on Bluetooth and Bluetooth Low
  Energy}.
\newblock {\em Transactions on Privacy and Security (TOPS)}, 2020.

\bibitem{ctkd-fluoride}
AOSP.
\newblock {Fluoride Bluetooth stack}.
\newblock
  \url{https://chromium.googlesource.com/aosp/platform/system/bt/+/master/README.md},
  Accessed: 2020-01-27, 2020.

\bibitem{pyca-crypto}
Python~Cryptographic Authority.
\newblock Python cryptography.
\newblock \url{https://cryptography.io/en/latest/}, Accessed: 2019-02-04, 2019.

\bibitem{bai2016staying}
Xiaolong Bai, Luyi Xing, Nan Zhang, XiaoFeng Wang, Xiaojing Liao, Tongxin Li,
  and Shi-Min Hu.
\newblock Staying secure and unprepared: Understanding and mitigating the
  security risks of apple zeroconf.
\newblock In {\em 2016 IEEE Symposium on Security and Privacy (SP)}, pages
  655--674. IEEE, 2016.

\bibitem{biham2018breaking}
Eli Biham and Lior Neumann.
\newblock Breaking the bluetooth pairing--fixed coordinate invalid curve
  attack.
\newblock
  \url{http://www.cs.technion.ac.il/~biham/BT/bt-fixed-coordinate-invalid-curve-attack.pdf},
  2018.

\bibitem{btsig-52}
{Bluetooth SIG}.
\newblock {Bluetooth Core Specification v5.2}.
\newblock
  \url{https://www.bluetooth.org/docman/handlers/downloaddoc.ashx?doc_id=478726},
  Accessed: 2020-01-27, 2019.

\bibitem{btsig-markets}
{Bluetooth SIG}.
\newblock {Bluetooth Markets}.
\newblock \url{https://www.bluetooth.com/markets/}, 2019.

\bibitem{btsig-market20}
{Bluetooth SIG}.
\newblock {Bluetooth Market Update 2020}.
\newblock \url{https://www.bluetooth.com/bluetooth-resources/2020-bmu/}, 2020.

\bibitem{ctkd-bluez}
BlueZ.
\newblock {Bluetooth 4.2 features going to the 3.19 kernel release}.
\newblock \url{https://tinyurl.com/q9dzh2h}, Accessed: 2020-01-27, 2014.

\bibitem{celosia2019fingerprinting}
Guillaume Celosia and Mathieu Cunche.
\newblock Fingerprinting bluetooth-low-energy devices based on the generic
  attribute profile.
\newblock In {\em Proceedings of the 2nd International ACM Workshop on Security
  and Privacy for the Internet-of-Things}, pages 24--31, 2019.

\bibitem{ctkd-cypress}
Cypress.
\newblock {BLE and Bluetooth}.
\newblock \url{https://www.cypress.com/products/ble-bluetooth}, Accessed:
  2020-01-27, 2019.

\bibitem{cyw920819evb-02}
Cypress.
\newblock {CYW920819EVB-02 Evaluation Kit}.
\newblock
  \url{https://www.cypress.com/documentation/development-kitsboards/cyw920819evb-02-evaluation-kit},
  Accessed: 2019-11-16, 2019.

\bibitem{dunning2010taming}
John Dunning.
\newblock Taming the blue beast: A survey of bluetooth based threats.
\newblock {\em IEEE Security \& Privacy}, 8(2):20--27, 2010.

\bibitem{fluhrer2001analysis}
Scott Fluhrer and Stefan Lucks.
\newblock {Analysis of the E0 encryption system}.
\newblock In {\em Proceedings of the International Workshop on Selected Areas
  in Cryptography}, pages 38--48. Springer, 2001.

\bibitem{garbelini2020sweyntooth}
{Garbelini, Matheus and Chattopadhyay, Sudipta and Wang, Chundong}.
\newblock {SweynTooth: Unleashing Mayhem over Bluetooth Low Energy}.
\newblock
  \url{https://asset-group.github.io/disclosures/sweyntooth/sweyntooth.pdf},
  Accessed: 2020-04-08, 2020.

\bibitem{haartsen1998bluetooth}
Jaap Haartsen, Mahmoud Naghshineh, Jon Inouye, Olaf~J Joeressen, and Warren
  Allen.
\newblock Bluetooth: Vision, goals, and architecture.
\newblock {\em ACM SIGMOBILE Mobile Computing and Communications Review},
  2(4):38--45, 1998.

\bibitem{haataja2010two}
Keijo Haataja and Pekka Toivanen.
\newblock Two practical man-in-the-middle attacks on {Bluetooth} secure simple
  pairing and countermeasures.
\newblock {\em Transactions on Wireless Communications}, 9(1):384--392, 2010.

\bibitem{hypponen2007nino}
Konstantin Hypponen and Keijo~MJ Haataja.
\newblock ``nino'' man-in-the-middle attack on bluetooth secure simple pairing.
\newblock In {\em Proceedings of the International Conference in Central Asia
  on Internet}, pages 1--5. IEEE, 2007.

\bibitem{ctkd-intel}
Intel.
\newblock {Intel Wireless Solutions}.
\newblock \url{https://www.intel.com/content/www/us/en/products/wireless.html},
  Accessed: 2020-01-27, 2019.

\bibitem{jakobsson2001security}
Markus Jakobsson and Susanne Wetzel.
\newblock Security weaknesses in {B}luetooth.
\newblock In {\em Proceedings of the Cryptographers' Track at the RSA
  Conference}, pages 176--191. Springer, 2001.

\bibitem{jasek2016gattacking}
S{\l}awomir Jasek.
\newblock Gattacking bluetooth smart devices.
\newblock Black Hat USA Conference, 2016.

\bibitem{jonsson2002security}
Jakob Jonsson.
\newblock On the security of {CTR}+ {CBC}-{MAC}.
\newblock In {\em Proceedings of the International Workshop on Selected Areas
  in Cryptography}, pages 76--93. Springer, 2002.

\bibitem{kelsey1999key}
John Kelsey, Bruce Schneier, and David Wagner.
\newblock Key schedule weaknesses in {SAFER+}.
\newblock In {\em Proceedings of the Advanced Encryption Standard Candidate
  Conference}, pages 155--167. NIST, 1999.

\bibitem{levi2004relay}
Albert Levi, Erhan {\c{C}}etinta{\c{s}}, Murat Aydos, {\c{C}}etin~Kaya
  Ko{\c{c}}, and M~Ufuk {\c{C}}a{\u{g}}layan.
\newblock Relay attacks on {Bluetooth} authentication and solutions.
\newblock In {\em Proceedings International Symposium on Computer and
  Information Sciences}, pages 278--288. Springer, 2004.

\bibitem{lindell2008attacks}
Andrew~Y Lindell.
\newblock Attacks on the pairing protocol of {Bluetooth} v2.1.
\newblock {\em Black Hat USA, Las Vegas, Nevada}, 2008.

\bibitem{internalblue}
Dennis Mantz, Jiska Classen, Matthias Schulz, and Matthias Hollick.
\newblock {InternalBlue} - {Bluetooth} binary patching and experimentation
  framework.
\newblock In {\em Proceedings of Conference on Mobile Systems, Applications and
  Services (MobiSys)}. ACM, June 2019.

\bibitem{minar2012bluetooth}
Nateq Be-Nazir~Ibn Minar and Mohammed Tarique.
\newblock Bluetooth security threats and solutions: a survey.
\newblock {\em International Journal of Distributed and Parallel Systems},
  3(1):127, 2012.

\bibitem{ubertooth}
Michael Ossmann.
\newblock Project {U}bertooth.
\newblock \url{https://github.com/greatscottgadgets/ubertooth}, Accessed:
  2019-10-21, 2019.

\bibitem{padgette2017guide}
John Padgette.
\newblock Guide to bluetooth security.
\newblock {\em NIST Special Publication}, 800:121, 2017.

\bibitem{ctkd-qualcomm}
Qualcomm.
\newblock {Expand the potential of Bluetooth}.
\newblock \url{https://www.qualcomm.com/products/bluetooth}, Accessed:
  2020-01-27, 2019.

\bibitem{rogaway2011evaluation}
Phillip Rogaway.
\newblock Evaluation of some blockcipher modes of operation.
\newblock {\em Cryptography Research and Evaluation Committees (CRYPTREC) for
  the Government of Japan}, 2011.

\bibitem{ryan2013bluetooth}
Mike Ryan.
\newblock Bluetooth: With low energy comes low security.
\newblock In {\em Proceedings of USENIX Workshop on Offensive Technologies
  (WOOT)}, volume~13, pages 4--4. USENIX, 2013.

\bibitem{armis2017blueborn}
Ben Seri and Gregory Vishnepolsky.
\newblock {The Attack Vector BlueBorne Exposes Almost Every Connected Device}.
\newblock \url{https://armis.com/blueborne/}, Accessed: 2018-01-26, 2017.

\bibitem{armis2019bleedingbit}
Ben Seri, Gregory Vishnepolsky, and Dor Zusman.
\newblock {BLEEDINGBIT}: The hidden attack surface within {BLE} chips.
\newblock \url{https://armis.com/bleedingbit/}, Accessed: 2019-07-24, 2019.

\bibitem{shaked2005cracking}
Yaniv Shaked and Avishai Wool.
\newblock Cracking the {B}luetooth {PIN}.
\newblock In {\em Proceedings of the conference on Mobile systems,
  applications, and services (MobiSys)}, pages 39--50. ACM, 2005.

\bibitem{sun2018man}
Da-Zhi Sun, Yi~Mu, and Willy Susilo.
\newblock Man-in-the-middle attacks on {Secure Simple Pairing} in {Bluetooth}
  standard v5. 0 and its countermeasure.
\newblock {\em Personal and Ubiquitous Computing}, 22(1):55--67, 2018.

\bibitem{von21method}
Maximilian von Tschirschnitz, Ludwig Peuckert, Fabian Franzen, and Jens
  Grossklags.
\newblock {Method Confusion Attack on Bluetooth Pairing}.
\newblock In {\em Proceedings of Symposium on Security and Privacy (S\&P)}.
  IEEE, 2021.

\bibitem{wong2005repairing}
Ford-Long Wong, Frank Stajano, and Jolyon Clulow.
\newblock Repairing the {Bluetooth} pairing protocol.
\newblock In {\em Proceedings of International Workshop on Security Protocols},
  pages 31--45. Springer, 2005.

\bibitem{bt-miscon2}
Joshua Wright.
\newblock {I Can Hear You Now - Eavesdropping on Bluetooth Headsets}.
\newblock
  \url{https://www.willhackforsushi.com/presentations/icanhearyounow-sansns2007.pdf},
  Accessed: 2018-10-30, 2018.

\bibitem{wu20blesa}
Jianliang Wu, Yuhong Nan, Vireshwar Kumar, Dave~Jing Tian, Antonio Bianchi,
  Mathias Payer, and Dongyan Xu.
\newblock {BLESA: Spoofing Attacks against Reconnections in Bluetooth Low
  Energy}.
\newblock In {\em 14th {USENIX} Workshop on Offensive Technologies ({WOOT})},
  2020.

\bibitem{ctkd-apple}
Apple WWDC.
\newblock {What's New in Core Bluetooth}.
\newblock \url{https://developer.apple.com/videos/play/wwdc2019/901}, Accessed:
  2020-01-27, 2019.

\bibitem{zhang2020breaking}
Yue Zhang, Jian Weng, Rajib Dey, Yier Jin, Zhiqiang Lin, and Xinwen Fu.
\newblock {Breaking Secure Pairing of Bluetooth Low Energy Using Downgrade
  Attacks}.
\newblock In {\em 29th {USENIX} Security Symposium ({USENIX} Security 20)},
  pages 37--54, 2020.

\end{thebibliography}





\end{document}